\documentclass[preprint,times]{elsarticle}

\usepackage{amssymb}
\usepackage{subcaption}
\usepackage{float}
\usepackage[3Dmenu]{media9}
\usepackage{hyperref}
\usepackage[UKenglish]{babel}
\usepackage{xcolor}
\usepackage{bm}
\usepackage{rsfso}
\usepackage{amsmath}

\begin{document}

\begin{frontmatter}

%% Title, authors and addresses

%% use the tnoteref command within \title for footnotes;
%% use the tnotetext command for theassociated footnote;
%% use the fnref command within \author or \address for footnotes;
%% use the fntext command for theassociated footnote;
%% use the corref command within \author for corresponding author footnotes;
%% use the cortext command for theassociated footnote;
%% use the ead command for the email address,
%% and the form \ead[url] for the home page:
%% \title{Title\tnoteref{label1}}
%% \tnotetext[label1]{}
%% \author{Name\corref{cor1}\fnref{label2}}
%% \ead{email address}
%% \ead[url]{home page}
%% \fntext[label2]{}
%% \cortext[cor1]{}
%% \affiliation{organization={},
%%             addressline={},
%%             city={},
%%             postcode={},
%%             state={},
%%             country={}}
%% \fntext[label3]{}

\title{Improved Velocity-Verlet Algorithm for the Discrete Element Method}
%\title{We need a title that more directly points out the sticking issue}

\author[1]{Dhairya R. Vyas}
\author[1,2,3]{Julio M. Ottino}
\author[1,2,3]{Richard M. Lueptow}
\author[1]{Paul B. Umbanhowar}

% \contrib[\authfn{1}]{Equally contributing authors.}

% Include full affiliation details for all authors
\affiliation[1]{Department of Mechanical Engineering, Northwestern University, Evanston, Illinois 60208, USA.}
\affiliation[2]{Department of Chemical and Biological Engineering, Northwestern University, Evanston, Illinois 60208, USA.}
\affiliation[3]{Northwestern Institute on Complex Systems (NICO), Northwestern University, Evanston, Illinois 60208, USA.}

%% use optional labels to link authors explicitly to addresses:
%% \author[label1,label2]{}
%% \affiliation[label1]{organization={},
%%             addressline={},
%%             city={},
%%             postcode={},
%%             state={},
%%             country={}}
%%
%% \affiliation[label2]{organization={},
%%             addressline={},
%%             city={},
%%             postcode={},
%%             state={},
%%             country={}}

% \author{}
% \affiliation{organization={},%Department and Organization
%             addressline={}, 
%             city={},
%             postcode={}, 
%             state={},
%             country={}}

\begin{abstract}
%% Text of abstract
The Discrete Element Method is widely employed for simulating granular flows, but conventional integration techniques may produce unphysical results for simulations with static friction when particle size ratios exceed $\mathcal{R} \approx 3$. These inaccuracies arise because some variables in the velocity-Verlet algorithm are calculated at the half-timestep, while others are computed at the full timestep. To correct this, we develop an improved velocity-Verlet integration algorithm to ensure physically accurate outcomes up to the largest size ratios examined ($\mathcal{R}=100$).
The implementation of this improved integration method within the LAMMPS framework is detailed, and its effectiveness is validated through a simple three-particle test case and a more general example of granular flow in mixtures with large size-ratios, for which we provide general guidelines for selecting simulation parameters and accurately modeling inelasticity in large particle size-ratio simulations.
\end{abstract}

%%Graphical abstract
% \begin{graphicalabstract}
%\includegraphics{grabs}
% \end{graphicalabstract}

%%Research highlights
% \begin{highlights}
% \item Research highlight 1
% \item Research highlight 2
% \end{highlights}


\end{frontmatter}

%% main text
\section{Introduction}
\label{}

Simulation codes and the models and algorithms buried deep within them can sometimes generate unphysical or conflicting results.  A famous recent case is the seven-year-long controversy over supercooled water, where slight differences in molecular dynamics codes resulted in strikingly different results \cite{Smart2018}.  Here we describe and remedy a similar subtle problem in several open-source Discrete Element Method (DEM) codes used for analyzing granular flows.

Granular materials, which include particles ranging in size from micrometers to meters, are the second most used substances in industry after water~\cite{duranIntroduction2000} and are encountered in a variety of industrial sectors such as pharmaceuticals~\cite{sarkarRoleForcesGoverning2017}, construction materials~\cite{johnParticleBreakageConstruction2023}, food processing~\cite{horabikParametersContactModels2016} and chemical processing~\cite{floreAspectsGranulationChemical2009}. Granular materials are also integral to natural and geological phenomena like avalanches, landslides, and asteroid impacts~\cite{meloshMechanicsLargeRock1987,grayRapidGranularAvalanches2003, sanchezSIMULATINGASTEROIDRUBBLE2011}. In many practical applications involving granular materials, particle size often varies widely; for example, uncrushed rock in mining operations is thousands of times larger than the resulting crushed fine powder~\cite{shinEffectBallSize2013}, soil particles may range from coarse sand to fine clay~\cite{hattoriRockFragmentationParticle1999}, and debris from avalanches and asteroids can vary greatly in scale~\cite{crostaFragmentationValPola2007}. 
Likewise, many processes in the chemical and pharmaceutical industries involve particles with sizes spanning two or more orders of magnitude~\cite{shekunovParticleSizeAnalysis2007,tongNumericalStudyEffects2010}.
Accurately accounting for these size variations is crucial for research related to these applications.

One of the most commonly used numerical methods for simulating granular flow applications is the Discrete Element Method (DEM)~\cite{cundallDiscreteNumericalModel1979b,osullivanParticulateDiscreteElement2014}. In DEM, which considers `soft' particles, local particle deformations are assumed to be small and are expressed as overlaps between adjacent contacting particles. Overlaps are used to calculate the contact forces, which determine the subsequent changes in velocity and configuration of the particles.

Many open-source DEM codes exist for simulating granular systems. Among the most widely used \cite{dostaComparingOpensourceFrameworks2024} are LAMMPS~\cite{thompsonLAMMPSFlexibleSimulation2022}, LIGGGHTS~\cite{klossModelsAlgorithmsValidation2012}, YADE~\cite{smilauerYadeDocumentation2021}, MercuryDPM~\cite{weinhartFastFlexibleParticle2020}, MUSEN~\cite{dostaMUSENOpensourceFramework2020}, GranOO~\cite{ANDRE201440} and MFIX~\cite{syamlalMFIXDocumentationTheory1993}. These solvers have been adopted to simulate granular flows in various applications such as hoppers~\cite{langstonDiscreteElementSimulation1995a, holzingerEffectChuteStart2022}, mills~\cite{mishraDiscreteElementMethod1992}, mixers~\cite{kanekoNumericalAnalysisParticle2000}, drums~\cite{kumarEffectAspectRatio2023}, and granular compaction~\cite{jerierStudyColdPowder2011}, as well as in geophysical and astrophysical applications such as avalanches~\cite{salciariniDiscreteElementModeling2010a}, volcanoes~\cite{morganDiscreteElementSimulations2005} and earthquakes~\cite{garciaDiscreteElementAnalysis2022}. Most of these specific simulation examples use relatively small ranges of particle sizes, with typical large to small particle size ratios $\mathcal{R}<3$, primarily because of the computational burden associated with larger size ratios. This limited size ratio range stands in stark contrast to the wide range of particle sizes observed in real-world granular systems.
%
% An alternate approach to soft particle DEM simulations is hard-particle event-driven simulations~\cite{campbellComputerSimulationGranular1985}. While this latter approach can be particularly useful for simple dilute granular systems~\cite{lomineTransportSmallParticles2006, liSpontaneousInterparticlePercolation2010c,roozbahaniMechanicalTrappingFine2014a, kerimovMechanicalTrappingParticles2018b}, its practicality for more realistic applications is limited due to its inherent difficulties in accurately modeling dense, inelastic flows where collision frequencies can diverge, resulting in simulations becoming impeded due to so-called inelastic collapse~\cite{bergerChallengesIIWide2014, ketterhagenStressResultsTwodimensional2005, deenReviewDiscreteParticle2007, mullerObliqueImpactFrictionless2012}.
 
One of the primary challenges in DEM simulations involving polydisperse systems is the rapid increase in computational resources needed as the size ratio increases. Berger et al.~\cite{bergerChallengesIIWide2014} highlight the difficulties of simulating highly polydisperse systems, including the need for shorter timesteps to accommodate interactions between particles with large size differences, and the increased complexity in neighbor detection and parallelization. However, over the past decade, advances in computational power and algorithms have facilitated more robust simulations that make possible considering physically relevant systems with size ratios $\mathcal{R}>3$. For instance, Remond~\cite{remondSimulationSmallParticles2010a} successfully simulated static granular beds with $\mathcal{R}$ up to 12.5; Gao et al.~\cite{gaoPercolationFineParticle2023a,gaoVerticalVelocitySmall2024a} considered flowing particles in sheared beds with $\mathcal{R}$ reaching 10; Lominé and Oger~\cite{lomineDispersionParticlesSpontaneous2009c} leveraged linked list algorithms for improved neighbor identification, simulating fine particle percolation through static particles with $\mathcal{R}$ up to 20.

A breakthrough for simulating large size ratios was achieved with Ogarko and Luding’s~\cite{ogarkoFastMultilevelAlgorithm2012} implementation of the hierarchical grid algorithm for neighbor detection, which can be effectively parallelized and showed a 220-fold speed increase over the classical linked cell method for $\mathcal{R}=50$. The integration of this algorithm into open-source platforms such as LAMMPS~\cite{shireSimulationsPolydisperseMedia2021}, MercuryDPM~\cite{weinhartFastFlexibleParticle2020} and MUSEN~\cite{dostaMUSENOpensourceFramework2020} has propelled further studies, enabling simulations with extremely large size ratios of up to $\mathcal{R}=200$~\cite{montiLargescaleFrictionlessJamming2022,montiFractalDimensionsJammed2023}. These advances underscore the growing interest and capability in exploring and simulating granular systems with large size ratios.

When we started investigating fine particle percolation in static beds, we aimed to extend existing work~\cite{remondSimulationSmallParticles2010a,gaoPercolationFineParticle2023a,gaoVerticalVelocitySmall2024a,lomineDispersionParticlesSpontaneous2009c} to size ratios as high as $\mathcal{R} = 50$. During this process, we discovered that DEM modeling approaches developed for low particle size ratios produce unphysical results for large size ratios. 
For example, at $\mathcal{R} = 20$, most fine particles (fines) percolating through a static bed are expected to freely pass through the interstices between the larger bed particles. However, we observe that many fines become trapped between two larger particles, oscillating along circular paths around the line connecting the two large particles.
Further investigation shows that this unphysical pendular motion occurs at size ratios as small as $\mathcal{R} = 3$. Thus, traditional DEM modeling approaches, though effective for small size ratios, $\mathcal{R}<3$, can result in unphysical and erroneous outcomes at larger size ratios, especially in static beds with low fine particle concentrations. In this paper, we address these issues using the open-source package LAMMPS, which is chosen for its wide utilization, scalability and integration of the hierarchical grid algorithm.
However, similar unphysical behavior also occurs with other standard DEM codes.

The primary focus of this paper is to delineate the nuances of the integration scheme used in DEM simulations, particularly focusing on its impact on the accuracy of the tangential static frictional force. Through a simple three-particle problem, we demonstrate how errors related to the velocity-Verlet integration scheme in current versions of widely-used open-source codes can lead to unphysical contact behavior, especially in systems with large size ratios. To address this problem, we propose an alternative implementation of the velocity-Verlet scheme that corrects these inaccuracies, and we validate this approach by comparing it with analytical results. Furthermore, we emphasize the importance of selecting appropriate contact stiffness values and timesteps for simulations involving large size ratios, providing guidelines for these selections. We also discuss the limitations of existing damping models in LAMMPS for accurately modeling inelasticity in large size ratio contacts and propose an alternative.  Several of these enhancements have been integrated into the latest developer version of LAMMPS, offering the community improved simulation accuracy for large size ratio systems. While the velocity-Verlet scheme is used as an example due to its implementation in most open-source codes including LAMMPS, LIGGGHTS, MercuryDPM, and GRANOO, the issues described here are relevant to other integration schemes as well.

This paper is organized as follows. Section 2  introduces a simple three-particle test case and highlights the unphysical behavior caused by the standard velocity-Verlet integration scheme. Section 3 explores the root cause of this issue in the velocity-Verlet method, and Section 4 presents an improved version  of the method that corrects these inaccuracies. In Section 5, the results from the improved scheme are compared with the standard approach using a particle percolation example, along with a discussion on selecting optimal spring stiffness, timesteps, and damping models for systems with large $\mathcal{R}$.

\section{Fine particle rolling between a large-particle pair}
\label{sec:stick}

\begin{figure}[h]
    \centering
    \begin{subfigure}[b]{0.4\textwidth}
        % \centering
        \includegraphics[width=\linewidth]{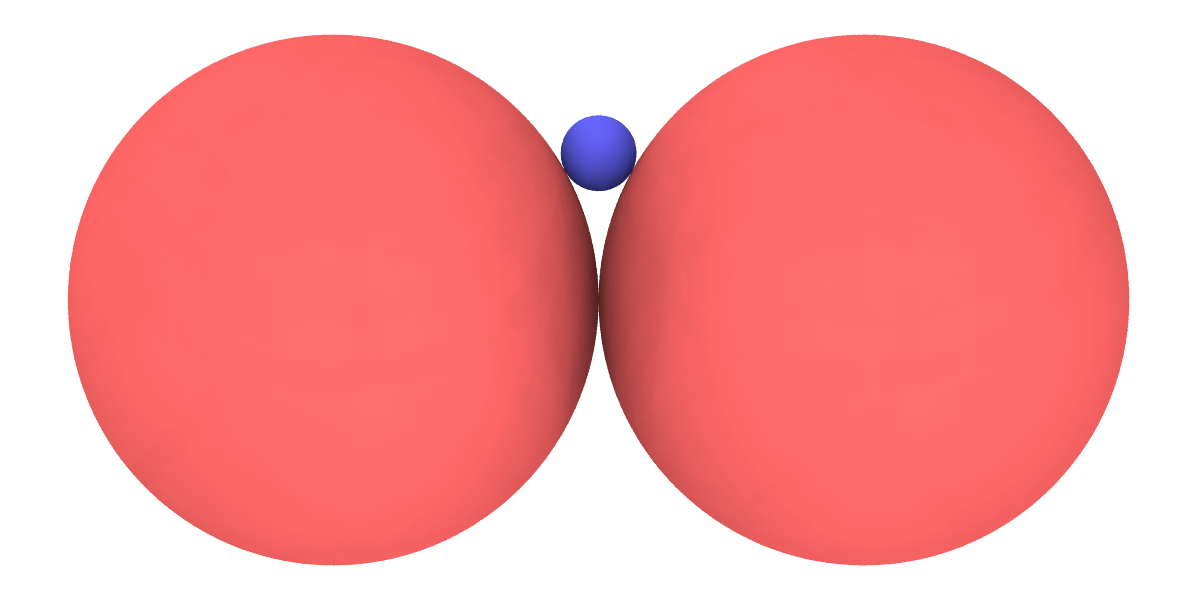}
        \caption{3D render (front view)}
    \end{subfigure}
    \begin{subfigure}[b]{\linewidth}
        \centering
        \includegraphics[width=0.5\linewidth]{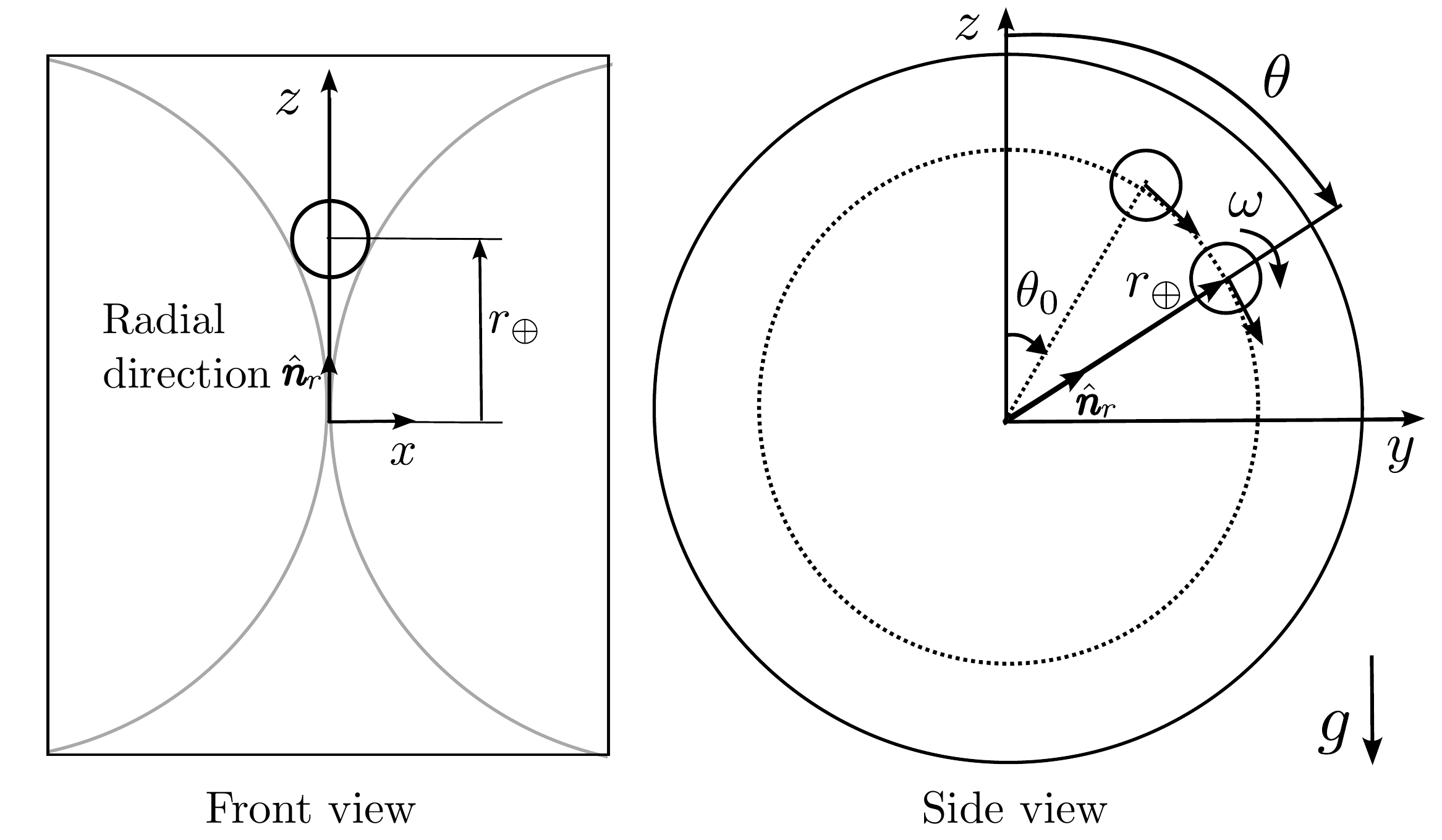}
        \caption{2D sketch}
    \end{subfigure}
    \caption{(a) 3D front view and (b) 2D sketch showing the front and side view of a spherical fine particle with diameter $2r$ released from rest at initial angle $\theta_0$ and rolling down the valley between two larger spherical particles with diameters $2R$ having their contact axis in the horizontal plane for size ratio $\mathcal{R}=7$. The distance of the fine particle from the line connecting the centers of the two large particles is $r_\oplus$ and the unit vector in the radial direction is $\hat{\bm{n}}_r$.}
    \label{fig:sch}
\end{figure}

To illustrate and understand the issue associated with the standard velocity-Verlet approach when simulating systems with large size ratios, we consider a simple three-body problem of a spherical fine particle rolling down the valley between a contacting pair of larger stationary spherical particles, see Figure~\ref{fig:sch}. The fine particle, which is initially tangent to both large particles,  is released from rest at at an initial angle $\theta_0$ with respect to vertical and moves along the valley between the two large particles under the influence of gravity, which acts vertically downward. 

\begin{figure}[h]
    \centering
    \includegraphics[width=0.5\linewidth]{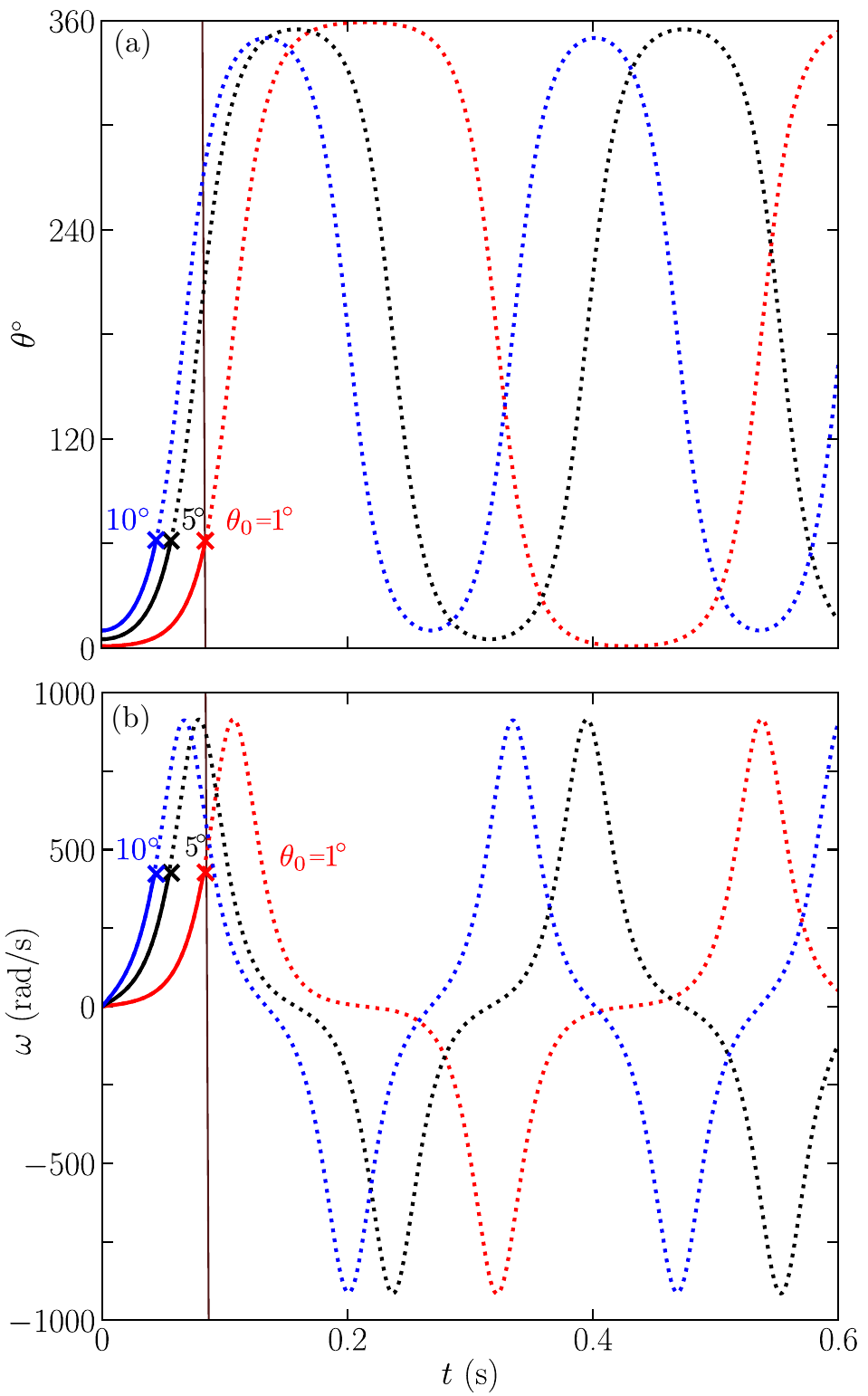}
    \caption{Comparison of DEM predictions (dotted curves) of (a) angular position, $\theta$ and (b) angular velocity, $\omega$,  with the analytic solutions (solid curves) for different initial fine particle positions,  $\theta_0$, for the three-particle case shown in Figures~\ref{fig:sch} and \ref{fig:anaschem} with $\mathcal{R} = 7$ and $\mu = 0.6$. Note that analytic solution for motion during contact ends at $\theta \approx 61^\circ$ when the fine particle separates from the large particles (denoted as $\times$) with the vertical line representing the instant when the fine particle separates from the large particles for $\theta_0=1^{\circ}$.}
    \label{fig:stick}
\end{figure}

An analytical solution to this problem, detailed in Appendix 1, is obtained by assuming that when a rigid fine particle is released at  angle $\theta_0$, it initially undergoes pure rolling motion while contacting both large particles. Once the tangential friction force $F_t$ reaches the sliding limit $\mu F_n$, where $\mu$ is the dynamic coefficient of friction and $F_n$ is the normal force, the particle begins to slide in addition to rolling, eventually losing contact with the two large particles and falling away. 
Consider, as an example, the time evolution of the position, $\theta$, and angular velocity, $\omega$, of the fine particle having diameter $2r$ in Figure~\ref{fig:sch}, with large particle diameters $2R = 4$\,mm, particle density $\rho=2500$\,kg/m$^3$, $\mu=0.6$, $g=9.81$ m/s$^2$, and a large-to-fine particle size ratio of $\mathcal{R} = R/r = 7$, shown as solid curves in Figure~\ref{fig:stick}. For all three values of $\theta_0$, the fine particle separates from the large particles at  $\theta \approx 61^\circ$, indicated by the $\times$ in Figure~\ref{fig:stick}(a). At this instant, the net radial force equals zero as the centrifugal acceleration exceeds the radial component of gravitational acceleration. For small $\theta_0$, it takes longer to build up the angular velocity, $\omega$, necessary for separation than for large $\theta_0$, but the value for $\omega$ when the fine particle detaches is similar in all three cases, shown as the solid curves in Figure~\ref{fig:stick}(b).

For comparison, DEM simulations of this problem are performed in LAMMPS using the \texttt{linear\_history} model with material properties of soda-lime glass~\cite{ashbyChapter15Material2013}: Young's modulus $E = 72$ GPa, Poisson ratio $\nu = 0.3$, and restitution coefficient $e_n = 0.8$. The diameter ($2R = 4$\,mm), density ($\rho=2500$\,kg/m$^3$), friction coefficient ($\mu=0.6$), and size ratio ($\mathcal{R} = 7$) are identical to those for the analytic solution. For this problem, assuming a characteristic velocity $V_0\sim\sqrt{2gR}$ = 0.198 m/s, the normal spring stiffness for the linear spring model is selected to match the Hertzian contact duration~\cite{thorntonInvestigationComparativeBehaviour2011c} using  
\begin{equation}
    k_n = 1.2024\left(m^{*^{1/2}} E^{*2} R^* V_0\right)^{2/5},
\end{equation}
where $m^*$ is the effective mass
\begin{equation}
    \frac{1}{m^*} = \frac{1}{m_{l}} + \frac{1}{m_{f}},
\end{equation}
$R^*$ is the effective radius
\begin{equation}
    \frac{1}{R^*} = \frac{1}{R} + \frac{1}{r},
\end{equation}
$ E^* $ is the effective Young's modulus
\begin{equation}
    \frac{1}{E^*} = \frac{(1-\nu_l^2)}{E_l}+\frac{(1-\nu_f^2)}{E_f},
\end{equation}
and the subscripts $l$ and $f$ denote large and fine particles, respectively. For these parameters values, $k_n =$ 3.26$\times$10$^5$\,N/m. 
The tangential spring stiffness is calculated using the approach suggested by Johnson~\cite{johnsonContactMechanics1985b}:
\begin{equation}
    \frac{k_t}{k_n} = \frac{(1-\nu_l)/G_l+((1-\nu_f)/G_f)}{(1-\nu_l/2)/G_l+(1-\nu_f/2)/G_f},
    \label{eq:johnson}
\end{equation}
where $G$ is the shear modulus. For our case, this reduces to~\cite{thorntonInvestigationComparativeBehaviour2011c}:
\begin{equation}
    \frac{k_t}{k_n} = \frac{2(1-\nu)}{2-\nu}.
    \label{eq:johnson2}
\end{equation}
For $\nu = 0.3$, $k_t = 0.8235k_n =$ 2.68$\times$10$^5$ N/m. To calculate the appropriate timestep, we first estimate the contact duration for a linear spring model: 
\begin{equation}
    t_c = \pi \sqrt{\frac{m^*}{k_n}}.
    \label{tc}
\end{equation}
For the particle properties in this example,  $t_c \approx 2.7$\,\textmu s.
Note that this applies to elastic contacts, which have shorter durations compared to inelastic contacts, making the elastic case the most restrictive. To ensure accurate discretization of the contact, the computational timestep should typically be less than $t_c/40$~\cite{thompsonGranularFlowFriction1991,fanShearRateIndependentDiffusionGranular2015a}. In this case, however, we choose a significantly smaller timestep of $t_c/270 = 10$~ns to guarantee more precise discretization. We do not simulate rolling or twisting friction in order to isolate the issue associated with the static friction term arising from the velocity-Verlet integration scheme.

\begin{figure}[h]
    \centering
    \includegraphics[width=0.5\linewidth]{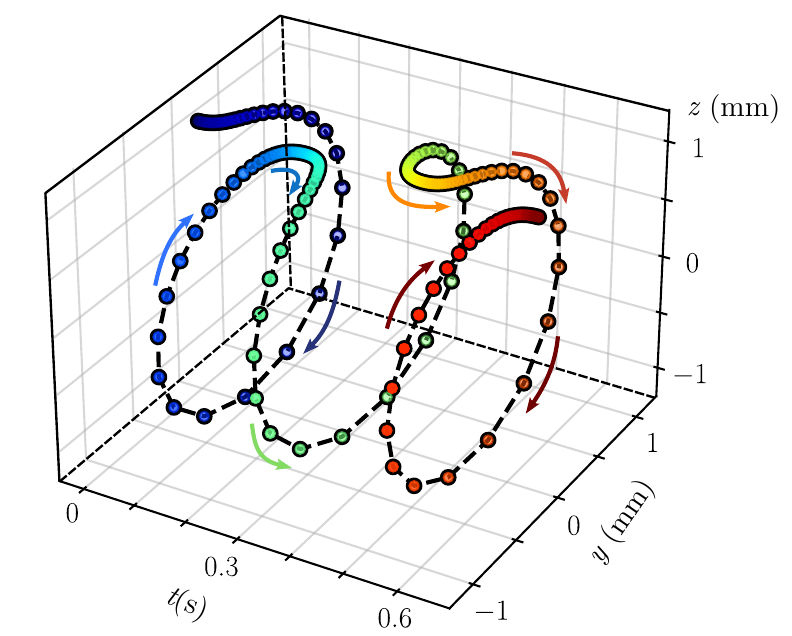}
    \caption{Temporal evolution of fine particle position in the $x =0$ plane for the conditions in Figure~\ref{fig:sch} with $\theta_0=1^{\circ}$, $\mathcal{R}=7$, and $\mu=0.6$.}
    \label{fig:3d2}
\end{figure}

Time series of $\theta$ and $\omega$ for three values of $\theta_0$ from simulation are plotted as dotted curves in Figure~\ref{fig:stick}.
Although the analytical and simulation results align well up to the point where the fine particle separates from the two large particles in the analytic solution (for $\theta \leq 61^\circ$ and $\omega \leq 425$ rad/s), the DEM simulation results continue after the fine particle should separate from the large particles, producing unphysical results that persist for the duration of the simulation. In other words, the DEM predictions indicate that after the fine particle starts at $\theta_0$ with $\omega = 0$, it rolls down the valley between the large particles (increasing $\theta$) with a corresponding increase in angular velocity $\omega$, but it never separates from the large particles. Instead, the fine particle continues to $\theta = 180^\circ$, where $\omega$ reaches its maximum. Subsequently, $\omega$ decreases to zero as $\theta$ reaches $-\theta_0$. The particle then reverses direction, rolling back to $\theta = \theta_0$, during which $\omega$ reaches a negative peak before returning to zero. This cycle repeats indefinitely, resulting in an unphysical pendular trajectory for the fine particle.
Figure~\ref{fig:3d2} shows this unphysical trajectory in the $x=0$ plane versus time using the coordinate system defined in Figure~\ref{fig:sch}, including the reversals in direction at $-\theta_0$ and subsequently at $\theta_0$ after the initiation of the motion (see video v1 in supplementary materials). 
Figure~\ref{fig:rr} shows the change $\Delta r_{\oplus}$ in radial distance $r_\oplus$ (shown in Figure~\ref{fig:sch}), of the fine particle compared to that for hard spherical particles, measured from the axis connecting the centers of the large particles to the center of the fine particle, during these oscillations. It starts with an initial negative value due to the initial overlap resulting from the `soft' DEM particles; however, the subsequent increase in the negative value indicates that with each pendular oscillation, the fine particle unphysically moves deeper into the valley between the large particles, rather than separating from them.

\begin{figure}[tbh]
    \centering
    \includegraphics[width=0.5\linewidth]{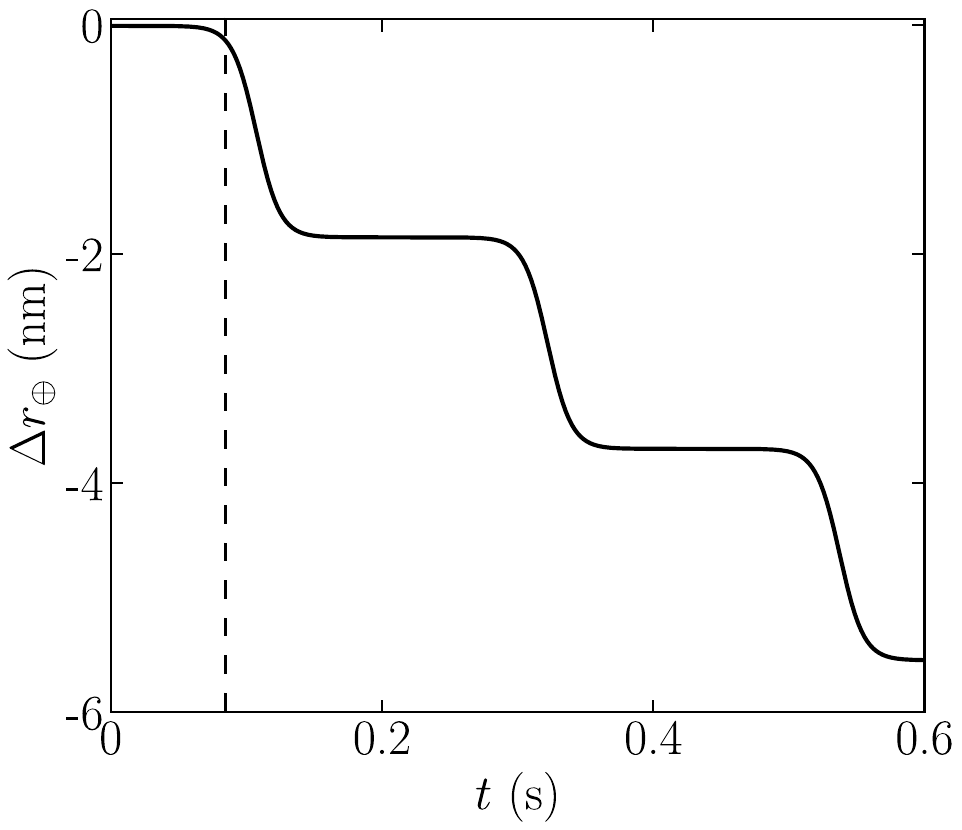}
    \caption{Deviation, $\Delta r_{\oplus}$, of the radial distance $r_{\oplus}$ of the fine particle from the line connecting the centers of the two large particles calculated using DEM relative to that for hard spherical particles for $\theta_0=1^{\circ}$, $\mathcal{R}=7$, and $\mu=0.6$.  The vertical dashed line marks the time when, according to the analytical solution, the fine particle should separate from the two large particles. The negative sign indicates that the fine particle is moving deeper into the valley between the two large particles with each pendular oscillation.}
    \label{fig:rr}
\end{figure}

\begin{figure}[h]
    \centering
    \includegraphics[width=0.5\linewidth]{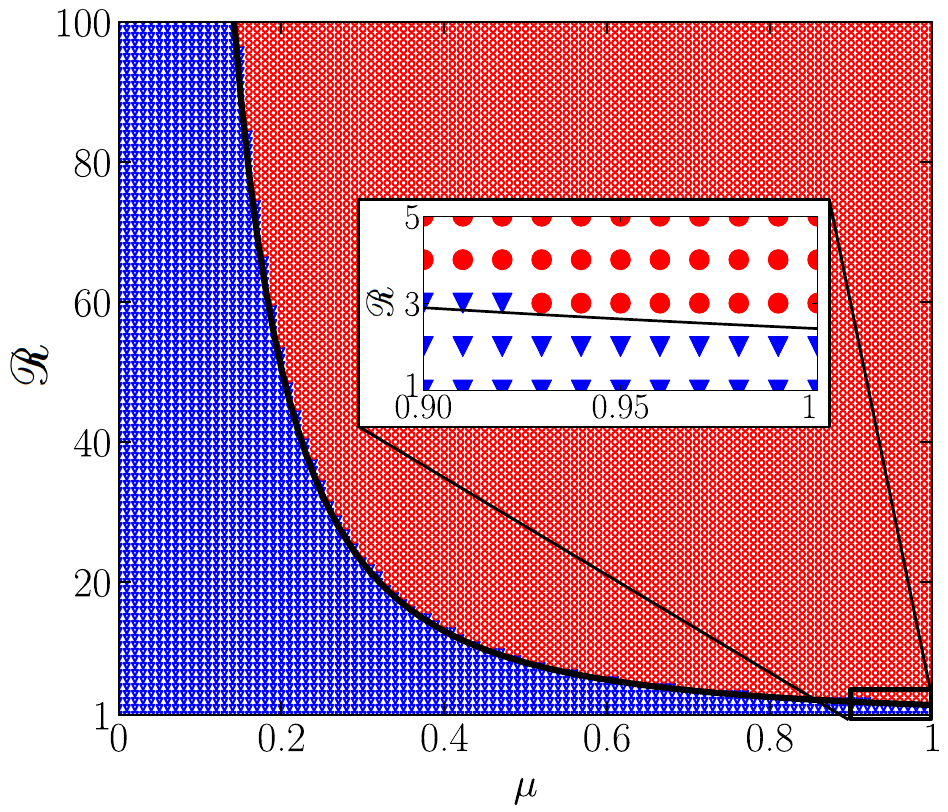}
    \caption{Phase diagram showing when the fine particle separates (blue) or unphysically remains in contact (red) with the two large particles vs. size ratio, $\mathcal{R}$, and friction coefficient, $\mu$, in DEM simulations for $\theta_0=5^{\circ}$. 10,100 data points are plotted on a grid with a spacing of 0.01 in $\mu$ and 1 in $\mathcal{R}$. Black curve indicates the analytical expression for $\mathcal{R}_c(\mu)$, Eq.~\ref{eq:psic}.} 
    \label{fig:musr}
\end{figure}

To more systematically investigate the effects of size ratio and friction coefficient, 10,100 DEM simulations were conducted following the same approach. The values of $\mathcal{R}$ were varied in increments of 1 and $\mu$ in increments of 0.01, tracking instances where the fine particle followed an unphysical trajectory and where it separates from the two large particles. As shown in Figure~\ref{fig:musr}, unphysical behavior is not confined to extreme size ratios but can occur for size ratios as small as $\mathcal{R}=3$ when $\mu$ is large (see inset). Because most prior DEM studies focused on monodisperse particles or small size ratios as well as moderate friction coefficients, this unphysical behavior was unlikely to significantly affect the DEM simulations results.  It is only as we consider wider ranges of particle sizes (large $\mathcal{R}$) that the unphysical results become evident. 

Unphysical behavior also occurs when rolling friction or a twisting force is included in modeling the problem shown in Figure~\ref{fig:sch}. For the spring-dashpot-slider (sds) rolling friction model~\cite{ludingCohesiveFrictionalPowders2008} used in LAMMPS, the fine particle exhibits pendular motion, but the oscillation amplitude decreases over time. In contrast, with Marshall's twisting friction model~\cite{MARSHALL20091541}, the fine particle rolls down and eventually stops. In neither case does the fine particle fall off.

It is clear that this behavior predicted by the $\texttt{linear\_history}$ model is incorrect. Similarly, the \texttt{mindlin} model in LAMMPS, the \texttt{hooke} and \texttt{hertz} models in LIGGGHTS, and the \texttt{LinearViscoelasticFrictionSpecies} model in MercuryDPM all produce identical incorrect results. A sensitivity analysis using LAMMPS was performed to investigate the influence of timestep size. In the current setup, using the LS model with $\mathcal{R} = 10$ and $\mu = 0.6$, the fine particle only separates from the pair of large particles when a very small timestep, less than $t_c/2000$, is used. When $\mu$ is increased to 0.9, this threshold decreases further to $t_c/20000$. However, using such extremely small timesteps substantially raises computational costs, and the $\mu$-dependence introduces uncertainty into the results.

\begin{figure}[tbh]
    \centering
    \includegraphics[width=0.5\linewidth]{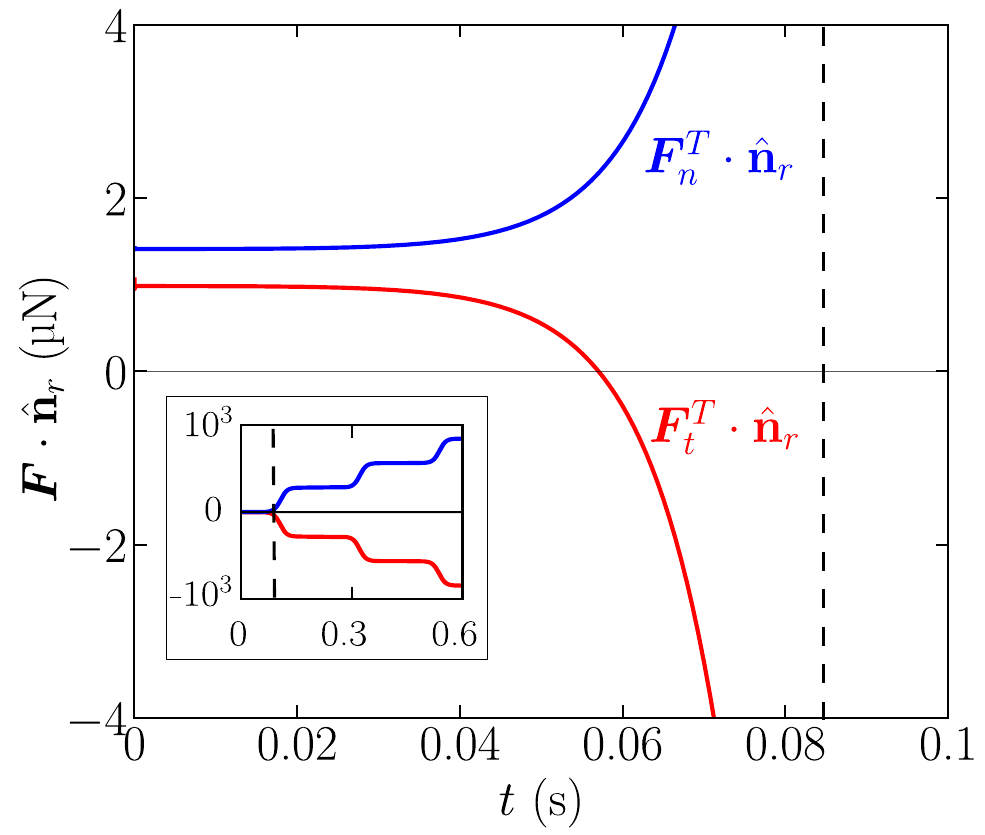}
    \caption{ Radial components of contact forces acting on the fine particle calculated using DEM for $\theta_0=1^{\circ}$, $\mathcal{R}=7$, and $\mu=0.6$. $\bm{F}_n^T$ is the total normal force, and $\bm{F}_t^T$ is the total tangential force. The vertical dashed line marks the time when, according to the analytical solution, contact should have broken. The inset shows that the contact forces continue to increase with each pendular oscillation and become several orders of magnitude larger than their initial values. For context, the weight of the fine particle is about 2.4~\textmu N.}
    \label{fig:ffine}
\end{figure}

The unphysical kinematics illustrated in Figures~\ref{fig:stick} to \ref{fig:musr} are associated with anomalies in the contact forces.
Figure~\ref{fig:ffine} shows the net radial components (along a line from the $x$-axis connecting the centers of the two large particles to the center of the fine particle, which is the $\hat{\bm{n}}_r$ direction in Figure~\ref{fig:sch}) of the normal, $\bm{F}_n$, and tangential, $\bm{F}_t$, contact forces acting on the fine particle as a function of time for the $\theta_0=1^\circ$ DEM simulation.
The superscript $T$ in $\bm{F}_n^T$ and $\bm{F}_t^T$ denotes the total normal and tangential forces (sum of the pairwise contact forces for both large particles) acting on the fine particle. The vertical dashed line marks the time when, according to the analytical solution, the fine particle should separate from the two large particles. 

At the onset of contact between the fine particle and the large particle, $\bm{F}_n^T$ and $\bm{F}_t^T$ are zero, but both change very quickly to become positive (which is not evident in Figure~\ref{fig:ffine} as it occurs within the first 10 \textmu s). This behavior is a result of the soft-sphere approach in DEM. When the fine particle is released between the two spheres at the initial angle $\theta_0$, the overlap is zero. However, as the fine particle moves downward due to gravity and begins to overlap with the large particles, a normal force is induced between the fine particle and each of the large particles. Concurrently, as the fine particle slides inward while overlapping, a tangential force also arises. Consequently, both the normal and tangential forces initially act in the radially outward direction from the centers of the large particles (positive $\hat{\bm{n}}_r$ direction) to counteract the radial component of the particle weight. From this time onward, $\bm{F}_n^T$ and $\bm{F}_t^T$ should both decrease as the radial component of the fine particle's weight decreases and the inertia related to centripetal acceleration associated with its circular trajectory between the two large particles increases.
What is critical, though, is that the DEM code instead calculates a sharp unphysical increase in the magnitude of these contact forces, driving $\bm{F}_n^T$ strongly outward and $\bm{F}_t^T$ strongly inward (see Figure~\ref{fig:ffine}).  
The increasing tangential force, $\bm{F}_t^T$, which acts radially inward, dominates, pushing the fine particle deeper into the valley between the two large particles, resulting in the unphysical behavior evident in Figure~\ref{fig:rr}. As a result of this incorrect calculation of $\bm{F}_n^T$ and $\bm{F}_t^T$, the fine particle does not separate from the two large particles at $\theta=61^\circ$, corresponding to the vertical dashed line at $t=0.84$~s in Figure~\ref{fig:ffine}. In fact, $\bm{F}_n^T$ and $\bm{F}_t^T$, as calculated by the DEM code, continue to increase, eventually reaching magnitudes several orders greater than their initial values, as shown in the inset of Figure~\ref{fig:ffine}. This is clearly unphysical, because $\bm{F}_n^T$ and $\bm{F}_t^T$ are forces of constraint determined by the particle weight and the initial position of the fine particle.

This erroneous behavior is especially puzzling, as the particle's inward radial motion should be countered by the tangential force $\bm{F}_t^T$, but, paradoxically, this force also acts radially inwards, exacerbating the error.  Instead of the fine particle rolling out of the valley between the two large particles, it is pushed deeper into the valley. The source of this anomaly lies in the velocity-Verlet integration scheme as described in detail in the next section.

\section{Standard velocity-Verlet integration scheme}
A DEM simulation begins by initializing the positions and velocities of particles and then models their subsequent development under the influence of contact and body forces. Various integration schemes are employed to numerically calculate this motion, with velocity-Verlet being particularly popular due to its balance of efficiency and second-order accuracy compared to simpler integrators such as the Euler method, which are only first-order accurate. The velocity-Verlet scheme requires only a single force calculation per timestep, unlike Runge-Kutta methods, which significantly reduces computational cost. Additionally, it has modest memory requirements, as it does not necessitate storing positions, velocities, and accelerations from previous timesteps, unlike Adams-Bashforth or higher-order back-difference methods. Crucially, the velocity-Verlet method is symplectic, meaning it preserves the geometric structure of Hamiltonian systems, which is necessary for conserving energy and momentum in simulations involving elastic, frictionless collisions. This balance of computational cost and accuracy makes the velocity-Verlet approach especially suitable for DEM problems where long-term stability and physical realism are essential \cite{thompsonLAMMPSFlexibleSimulation2022}. As a result, it has been incorporated into several open-source DEM packages including LAMMPS, LIGGGHTS, MercuryDPM, and GRANOO. 

\subsection{Standard DEM calculation}

\begin{figure}[H]
    \centering
    \includegraphics[width=0.5\linewidth]{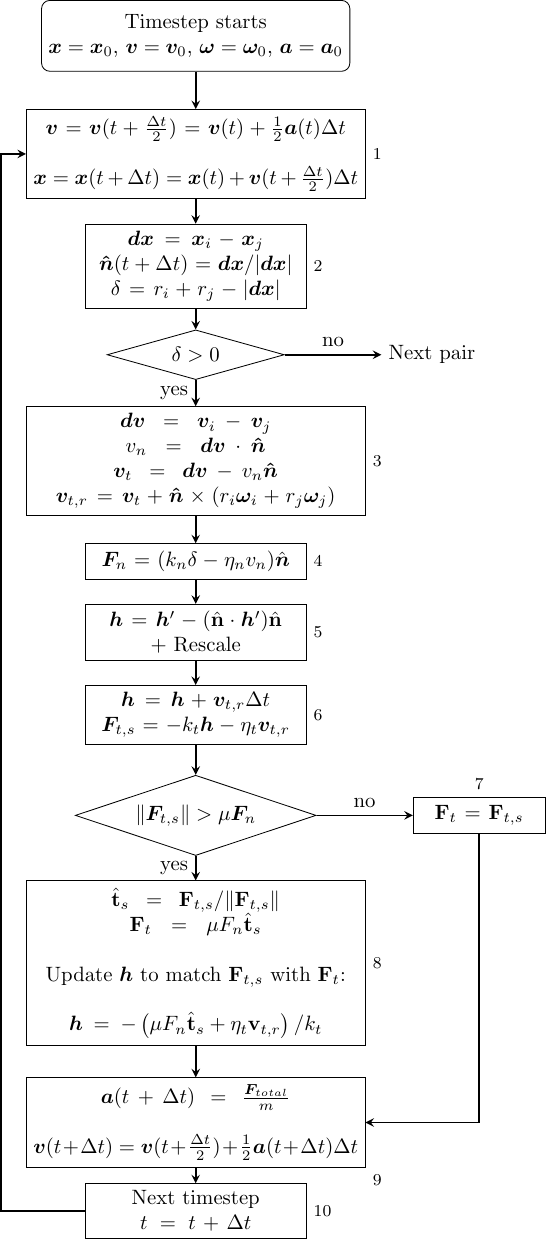}
    \caption{Flowchart for DEM with a linear spring contact model using standard velocity-Verlet approach.}
    \label{fig:flowchart}
\end{figure}

A flowchart of the DEM timestepping process using the velocity-Verlet method is shown in Figure~\ref{fig:flowchart}. Each timestep begins with the initial position $\bm{x}$ and velocity $\bm{v}$ of each spherical particle. If it is the first timestep, the acceleration $\bm{a}$ is initialized to $\bm{0}$, whereas in later timesteps $\bm{a}$ is calculated from the previous timestep. Half-step velocities are computed as
\begin{equation}
    \bm{v}\left(t+\frac {1}{2} \Delta t\right) = \bm{v}(t) + \frac {\Delta t}{2} \bm{a}(t),
\end{equation}
which are then used to determine new positions at $t+\Delta t$:
\begin{equation}
    \bm{x}\left(t+\Delta t\right) = \bm{x}(t) +  \bm{v}\left(t+\frac {1}{2} \Delta t\right) \Delta t.
\end{equation}
Pairwise overlaps are computed as
\begin{equation}
    \delta = r_i + r_j - |\bm{dx}|,
\end{equation}
where $r_i$ and $r_j$ are the radii of particles $i$ and $j$, and  $\bm{dx} = \bm{x}_i - \bm{x}_j$. If the pairwise overlap is positive, the particles are considered to be in contact, and the pairwise contact force is calculated using a contact model. The Linear Spring (LS) model and the Hertz-Mindlin model are the two most commonly used contact models for simulating granular contacts. Here, we utilize the LS model for its simplicity, though the discussion and findings are equally applicable to the Hertz-Mindlin model.

Before calculating the normal and tangential force components using the LS model, the relative velocity components between the contacting particles must be determined. The normal component is given by
\begin{equation}
    {v}_n = \bm{dv} \cdot \bm{\hat{n}},
\end{equation}
where $\bm{dv} = \bm{v}_i - \bm{v}_j$ and $\bm{\hat{n}} = \bm{dx} / |\bm{dx}|$, where the caret indicates a unit vector. The relative tangential velocity at the contact point is calculated by subtracting the normal component from $\bm{dv}$ and adding the tangential component induced by the angular velocity $\bm{\omega}$ of the contacting particles:
\begin{equation}
    \bm{v}_{t,r} = \bm{dv} - {v}_n \bm{\hat{n}} + \bm{\hat{n}} \times (r_i \bm{\omega}_i + r_j \bm{\omega}_j),
    \label{eq:vtr}
\end{equation}
where the first two terms represents the center of mass translational tangential velocity, $\bm{v}_t = \bm{dv} - v_n\bm{\hat{n}}$. The normal force between contacting particles, $\bm{F}_n$, is then determined using the LS model  and calculated as
\begin{equation}
    \bm{F}_n = (k_n \delta - \eta_n {v}_n)\hat{\bm{n}},
\end{equation}
where $k_n$ is the normal spring stiffness and $\eta_n$ is the normal damping coefficient, which depends on the restitution coefficient (details are discussed in Section~\ref{sec:guidelines}). 

For tangential forces, the static friction component is modeled using a tangential spring. The challenge lies in implementing a Hookean linear spring for the tangential force, reflecting static contact before sliding occurs. By definition, a spring model requires a non-zero deformation during the ``static'' friction phase. This deformation is based on the tangential contact history from the time when contact between two particles begins at $t_0$ to time $t$ within the current contact:
\begin{equation}
    \bm{h} = \int_{t_0}^t \mathbf{v}_{t,r}(\tau) \, \mathrm{d}\tau,
    \label{eq:history}
\end{equation}
where $\mathbf{v}_{t,r}$ is the relative tangential velocity at the contact point as defined in Eq.~\ref{eq:vtr}. This represents the elongation of the tangential spring, which might be thought of physically as shear deformation of the contacting particles relative to the point of contact before any sliding occurs.

In Step 6 of the flow chart in Figure~\ref{fig:flowchart}, the integral in Eq.~\ref{eq:history} is represented as a summation, where the value of $\bm{h}$ is updated by adding the incremental tangential displacement, $\mathbf{v}_{t,r}\Delta t$, at each timestep. As the local coordinate system for the contact evolves from $\bm{x}(t)$ to $\bm{x}(t+\Delta t)$, the tangential displacement history $\bm{h}$, before its update in Step 6, is rotated to account for this rotation of the particle pair reference frame \cite{ludingCohesiveFrictionalPowders2008}. The typical approach to rotate the accumulated displacement into the tangent plane removes any component of the net tangential displacement that lies along the current normal direction, $\hat{\mathbf{n}}$, and rescales $\bm{h}$ to preserve its magnitude:
\begin{equation}
    \bm{h} = \left(\bm{h}' - (\hat{\mathbf{n}} \cdot \bm{h}')\hat{\mathbf{n}}\right) \frac{|\bm{h}'|}{|\bm{h}' - (\hat{\mathbf{n}}\cdot\bm{h}')\hat{\mathbf{n}}|}.
    \label{eq:xiupdate}
\end{equation}
This is shown in Step 5 where $\bm{h}'$ is the accumulated displacement prior to the current timestep and $\bm{h}$ is the corrected displacement.

The tangential static frictional force, is calculated based on the corrected displacement as:
\begin{equation}
    \mathbf{F}_{t,s} = -k_t \bm{h}-\eta_t \mathbf{v}_{t,r},
    \label{eq:ls}
\end{equation}
where $\eta_t$ is the damping coefficient. As shown in the flowchart, when the magnitude of this tangential friction force exceeds Coulomb's sliding limit, $|\mathbf{F}_{t,s}| > \mu F_{n}$, where $F_n$ is the magnitude of the normal force, the tangential force is recalculated using Amontons' first law, which assumes that the tangential force is proportional to the normal force with the dynamic friction coefficient $\mu$ as the proportionality constant:
\begin{equation}
    \mathbf{F}_{t} = \mu F_{n} \hat{\mathbf{t}}_s,
    \label{eq:sliding}
\end{equation}
where $\hat{\mathbf{t}}_s$ is the unit vector in the direction of $\mathbf{F}_{t,s}$. To account for the pure sliding state of the contact, along with variations in the normal force and relative tangential velocity, the tangential history $\bm{h}$ is updated to match the sliding tangential force~\cite{ludingCohesiveFrictionalPowders2008}:
\begin{equation}
    \bm{h} = -\frac{1}{k_t}\left(\mu F_{n}\hat{\mathbf{t}}_s + \eta_t \mathbf{v}_{t,r}\right).
\end{equation}
This represents the maximum tangential spring elongation possible for an instantaneous value of normal force and relative tangential velocity. These pairwise normal and tangential forces, along with any other body forces, are then summed to get the total force acting on the particle, $\bm{F}_{\text{total}}$, which is then used to calculate acceleration at time $t + \Delta t$:
\begin{equation}
    \bm{a}(t + \Delta t) = \frac{\bm{F}_{\text{total}}}{m}.
\end{equation}
The acceleration is then used to update the half-step velocity to the full-step velocity, $\bm{v}(t+\Delta t)$, as 
\begin{equation}
    \bm{v}(t + \Delta t) = \bm{v}\left(t + \frac{\Delta t}{2}\right) + \frac{1}{2}\bm{a}(t + \Delta t) \Delta t. 
\end{equation}
Similar calculations update the torques and angular velocities of the particle pair. This is the standard velocity-Verlet integration scheme that is implemented in major open-source codes like LAMMPS, LIGGGHTS, and MercuryDPM at the time of publication.

\subsection{Source of the error}

In Figures~\ref{fig:rr} and \ref{fig:ffine}, we noted that the tangential force increases in the radially inward direction (toward the axis connecting the centers of the large particles) as the fine particle rolls down the valley between the two large particles, causing an inward motion that is physically unrealistic. From Step 6 of the flowchart in Figure~\ref{fig:flowchart}, the static tangential force depends on the history term $\bm{h}$, which is updated using Eq.~\ref{eq:xiupdate}. The history term relies on the relative tangential velocity $\bm{v}_{t,r}$ at the contact point, calculated in Step 3 using Eq.~\ref{eq:vtr}. In this calculation of $\bm{v}_{t,r}$, the normal vector $\hat{\bm{n}}(t + \Delta t)$ is derived from the particle positions at $t + \Delta t$ (Step 2), while the velocities used are from $t + \Delta t/2$ (Step 1). This causes the velocities to lag the positions by $\Delta t/2$.
This lag leads to an inaccurate projection of the velocity in the tangential direction, subsequently yielding an erroneous history term and ultimately resulting in an incorrect tangential force.

\begin{figure}[t]
    \centering
    \includegraphics[width=0.5\linewidth]{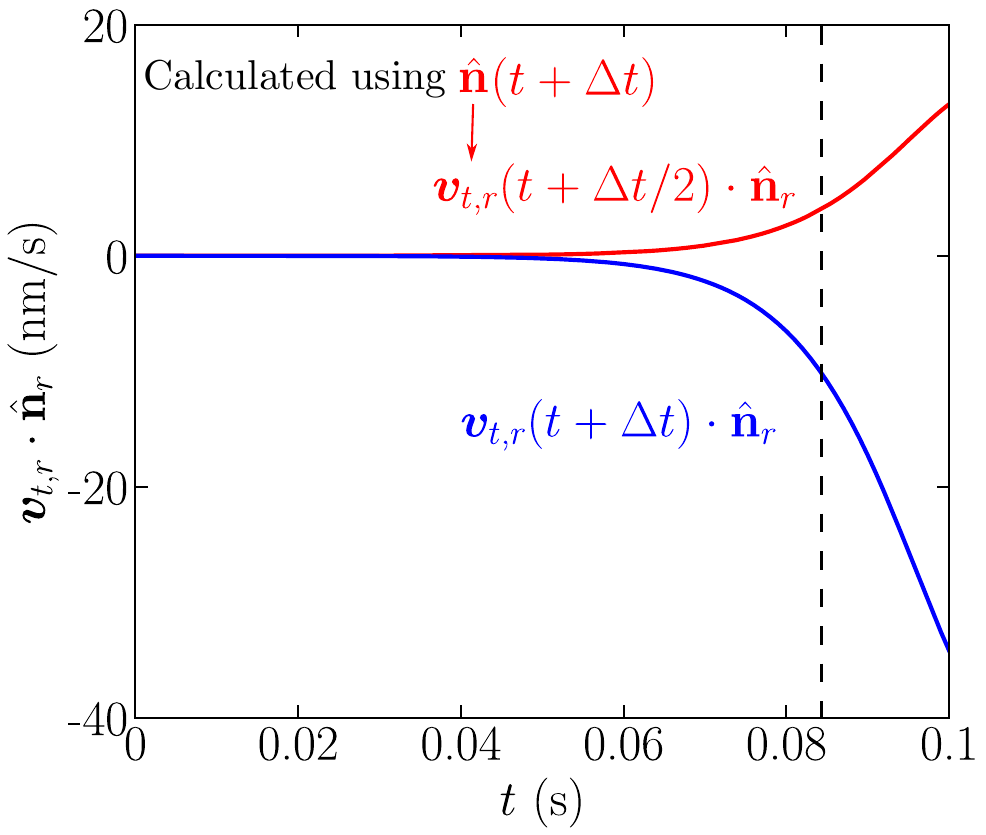}
    \caption{Comparison of the radial components of incorrect half-step tangential velocity $\bm{v}_{t,r}(t + \Delta t/2)$ and full-step tangential velocity $\bm{v}_{t,r}(t + \Delta t)$ for $\theta_0=1^{\circ}$, $\mathcal{R}=7$, and $\mu=0.6$. The vertical dashed line marks the time in the analytical solution when contact breaks.}
    \label{fig:vtr}
\end{figure}

We compare the radial components of $\bm{v}_{t,r}(t + \Delta t/2)$ and $\bm{v}_{t,r}(t + \Delta t)$ in Figure~\ref{fig:vtr}. Here, $\bm{v}_{t,r}(t + \Delta t/2)$ is calculated using the standard velocity-Verlet approach and includes a  lag of half a timestep, while $\bm{v}_{t,r}(t + \Delta t)$ is obtained using the positions from Step 1 and the velocities from Step 9, therefore both positions and velocities are at $t + \Delta t$. The half-step velocity components at $t + \Delta t/2$ are used to calculate the forces acting on the particles, while the full-step velocities at $t + \Delta t$ are used to update the particle positions.
As shown in Figure~\ref{fig:vtr}, the incorrect half-step tangential velocity, $\bm{v}_{t,r}(t + \Delta t/2)$, acts radially outward and results in an incorrect tangential force acting radially inward. Since the history term $\bm{h}$ is  the time integral of $\bm{v}_{t,r}(t + \Delta t/2)$, the error accumulates and $\bm{h}$ grows in the radially outward direction, resulting in a growing tangential force acting radially inward ($\bm{F}_t^T$ in Figure~\ref{fig:ffine}).
This incorrect tangential force is then used to calculate the full-step tangential velocity, $\bm{v}_{t,r}(t + \Delta t)$ in Step 9. Rather than being zero before the sliding limit and pointing radially outward after it, the full-step tangential velocity erroneously points radially inward, driving the fine particle deeper into the valley between the two large particles, ultimately preventing the fine particle from separating from them.

\subsection{A simple physical model}

\begin{figure}[h]
    \centering
    \includegraphics[width=0.5\linewidth]{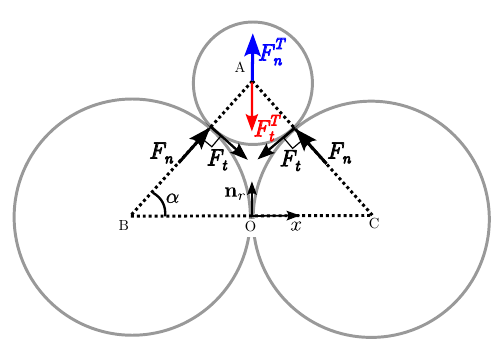}
    \caption{Schematic of opposing net normal and tangential forces acting on the fine particle.}
    \label{fig:mur}
\end{figure}

After analyzing the error in the standard velocity-Verlet integration scheme, we now explain the influence of $\mathcal{R}$ and $\mu$ on the fine particle behavior, as shown in Figure~\ref{fig:musr}. The question is: why does the error in the velocity-Verlet scheme lead to unphysical results for certain combinations of $\mathcal{R}$ and $\mu$, but not for others? The answer lies in the relative magnitudes of the net tangential contact force, $F_t^T$, and the net normal contact force, $F_n^T$, which act in the radial direction $\hat{\bm{n}}_r$, relative to the axis connecting the two large particles, as illustrated in Figure~\ref{fig:mur}. We assume that both large particles exert identical contact forces on the fine particle.
The net normal force always acts outward from the valley, with a magnitude of $F_n^T = 2 F_n \sin{\alpha}$, where $\alpha$ is defined in the figure and is determined by $\mathcal{R}$, i.e., $\cos{\alpha} = \mathcal{R}/(1+\mathcal{R})$ (see Eq.~\ref{eq:alpha} in Appendix 1).
In contrast, the magnitude of the net frictional force is bounded by $F_{t,max}^T = 2 \mu F_n \cos{\alpha}$ and can act in either the inward or outward direction.
We are concerned here with cases where the net frictional force opposes the net normal force. For these cases, there are two distinct regimes depending on their relative magnitudes. In the absence of other forces, when $F_n^T > F_{t,max}^T$, the fine particle will always slip out of the valley between the two large particles; when $F_n^T < F_{t,max}^T$, the fine particle will be trapped. The boundary between these two regimes is determined by equating $F_n^T$ and $F_{t,max}^T$, which gives 
\begin{equation}
    \mu = \tan{\alpha}.
    \label{eq:mutana}
\end{equation}
When $\mu > \tan{\alpha}$, the fine particle will be trapped, and when $\mu < \tan{\alpha}$, the fine particle will slip out.
Since $\alpha$ is determined by $\mathcal{R}$, an equivalent boundary between slipping and unphysical trapping can be expressed in terms of a critical size ratio, $\mathcal{R}_c$, and $\mu$ as
\begin{equation}
    \mathcal{R}_c = \frac{1}{\sqrt{1+\mu^2} - 1}.
    \label{eq:psic}
\end{equation}
Returning now to the results shown in Figure~\ref{fig:musr}, we see that the ($\mu$, $\mathcal{R}$) pairs associated with physical and unphysical behavior are well predicted by Eq.~\ref{eq:psic}. This occurs because, although the standard velocity-Verlet integration scheme always unphysically increases $F_t^T$ inward, it is only for $\mathcal{R} > \mathcal{R}_c$ that trapping can occur. When $\mathcal{R} < \mathcal{R}_c$, $F_n^T$ is always larger than $F_t^T$, and the fine particle always slips out eventually because the radial component of gravity decreases and the centripetal acceleration increases as the fine rolls down the valley.

\section{Improved velocity-Verlet integration}

The unphysical pendular motion described above is ultimately due to the $\Delta t/2$ phase difference between the positions and velocities during force calculations in the standard velocity-Verlet algorithm. To address this issue, one might consider alternative integration schemes where positions and velocities are updated simultaneously. However, the velocity-Verlet algorithm offers several advantages that make it preferable over other methods, as described earlier. 
Moreover, the velocity-Verlet algorithm is already widely implemented in standard open-source codes, and switching to a different integration scheme could require significant modifications to the codebase. Therefore, in this section, we present an improved version of the velocity-Verlet algorithm that addresses the issue while retaining its inherent advantages.

\begin{figure}[H]
    \centering
    \includegraphics[width=0.35\linewidth]{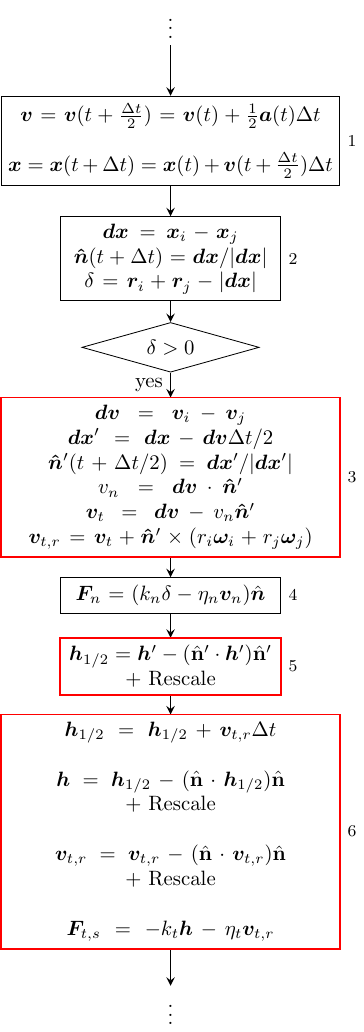}
    \caption{Flowchart showing modifications to the velocity-Verlet algorithm.}
    \label{fig:fc2}
\end{figure}

In Steps 3 to 9 of the flowchart for the standard velocity-Verlet approch in Figure~\ref{fig:flowchart}, the primary goal is to compute the acceleration $\bm{a}(t+\Delta t)$ and use it to update the velocity $\bm{v}(t+\Delta t)$. This requires the forces and torques at time $t + \Delta t$. Using Taylor series expansion, the history term $\bm{h}(t + \Delta t)$ can be expressed as:
\begin{equation}
    \bm{h}(t+\Delta t) = \bm{h}(t) + \frac{d\bm{h}}{dt}\Big|_{t} \Delta t.
\end{equation}
A central difference expansion of the derivative 
\begin{equation}
    \frac{d\bm{h}}{dt}\Big|_{t} = \frac{\bm{h}(t+\Delta t/2) - \bm{h}(t-\Delta t/2)}{2(\Delta t/2)} + O(\Delta t^2)
    \label{eq:history2}
\end{equation}
combined with the fact that the history term represents a displacement that can be updated using the velocity as 
\begin{equation}
    \bm{h}\left(t+\frac{\Delta t}{2}\right) = \bm{h}\left(t-\frac{\Delta t}{2}\right) + \bm{v}_{t,r}\left(t+\frac{\Delta t}{2}\right)  \Delta t,
\end{equation}
yields:
\begin{equation}
    \bm{h}(t+\Delta t) = \bm{h}(t) + \bm{v}_{t,r}\left(t+\frac{\Delta t}{2}\right) \Delta t.
    \label{eq:history3}
\end{equation}

This approximation allows the relative tangential velocity  at time  $t + \Delta t/2$ to be accurately calculated. The modifications to the velocity-Verlet algorithm to accommodate this are shown as a flowchart in Figure~\ref{fig:fc2}. 
In Step 3, before calculating the velocity components, we first compute the half-step normal vector $\bm{\hat{n}}'$ at $(t + \Delta t/2)$ using the updated position $\bm{x}(t + \Delta t)$ and velocity $\bm{v}(t + \Delta t/2)$. The half-step normal and tangential velocity components, $v_n(t + \Delta t/2)$ and $\bm{v}_{t,r}(t + \Delta t/2)$, are then determined using this $\bm{\hat{n}}'(t + \Delta t/2)$, rather than $\bm{\hat{n}}(t + \Delta t)$ in Step 2 of Figure~\ref{fig:flowchart}. Since both the normal vector and the velocities are evaluated at half-steps, there is no phase lag.

\begin{figure}[H]
    \centering
    \includegraphics[width=0.5\linewidth]{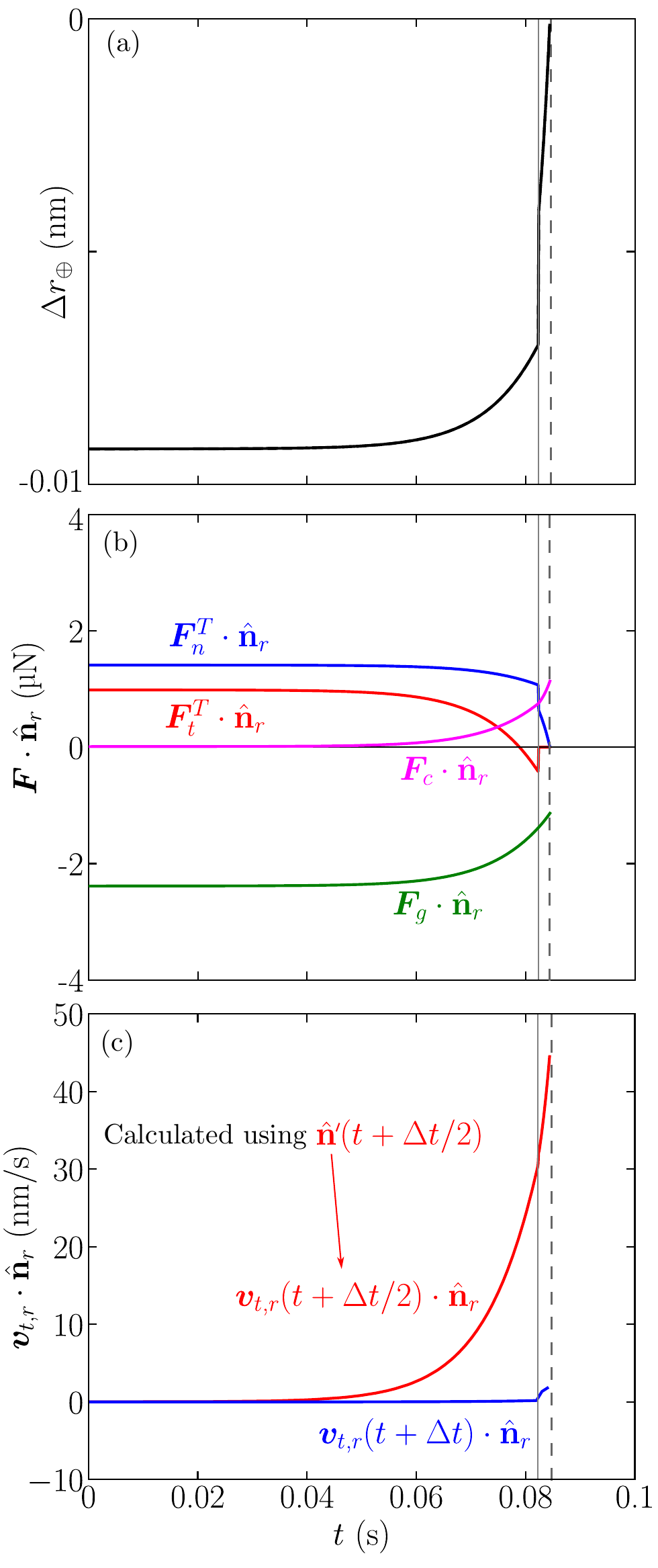}
    \caption{DEM simulation results for geometry in Figure~\ref{fig:sch} using improved velocity-Verlet integration scheme for $\theta_0=1^{\circ}$, $\mathcal{R}=7$, and $\mu=0.6$. (a) Deviation of the radial distance of particle to the axis between the large particles from that for hard particles, (b) radial components of forces acting on the fine particle, and (c) radial components of the half-step and full-step tangential velocity. Vertical solid line indicates when the sliding limit is reached by tangential spring and the vertical dashed line indicates  when contact breaks.}
    \label{fig:newf}
\end{figure}

Using the modified approach, the half-step normal velocity $v_n(t + \Delta t/2)$ is then used to calculate the damping term for the normal force in Step 4. For the tangential force in Step 6, the history displacement term from the previous timestep is first rotated to the half-step plane using $\bm{\hat{n}}'$ to obtain $\bm{h}_{1/2}$. This is then updated by adding the relative tangential velocity $\bm{v}_{t,r}(t + \Delta t/2)$, multiplied by timestep $\Delta t$. The resulting history term is then rotated from the half-step plane to the full-step plane using $\bm{\hat{n}}$ and rescaled. This updated history term is used to calculate the elastic component of the tangential force. The relative tangential velocity used in the damping term is also rotated and rescaled from the half-step plane to the full-step plane using $\bm{\hat{n}}$. These modifications ensure that the velocity components are in phase with the positions. The elastic normal force remains unaffected by these modifications, as it continues to use positions at $(t + \Delta t)$.

The influence of these modifications on the forces is shown in Figure~\ref{fig:newf}, which corresponds to Figures~\ref{fig:rr}, \ref{fig:ffine} and \ref{fig:vtr} for the unphysical case. 
In addition to the contact forces, the radial components of the gravitational force, $\bm{F}_g$, and the centrifugal force representing the inertia related to centripetal acceleration, $\bm{F}_c$, are also shown in Figure~\ref{fig:newf}(b).
At the onset of contact between the fine particle and the two large particles, the radial components of both the normal and tangential forces are positive and balance the radial component of the particle weight, $\bm{F}_g$. 
Unlike the unphysical case simulation results shown in Figure~\ref{fig:ffine}, $\bm{F}_n^T$ is now strictly decreasing.
As the fine particle rolls down the valley between the two large particles, the centrifugal force increases as it speeds up and the radial component of the gravitational force decreases. 
Due to this, the radial component of the half-step tangential velocity $\bm{v}_{t,r}(t + \Delta t/2)$ increases as indicated in Figure~\ref{fig:newf}(c). This half-step tangential velocity is then used to calculate the elastic component of the tangential force $\bm{F}_t$.
As the tangential spring associated with static friction stretches, the resulting tangential force opposes the particle's outward tangential motion, increasing in the radially inward direction. This maintains the circular motion of the fine particle about axis connecting the large particles, as indicated by the zero full-step radial component of the tangential velocity $\bm{v}_{t,r}(t + \Delta t)$. Once the static friction force exceeds the sliding limit (indicated by the vertical solid line), the tangential force transitions to that associated with dynamic friction. As the fine particle slips, both the normal and tangential forces quickly go to zero as the fine particle slides out of the valley losing contact with the large particles.  This is also evident in the deviation of the radial position of the fine particle with respect to the hard particle limit, $\Delta r_\oplus$, shown in Figure \ref{fig:newf}(a).  Initially, $\Delta r_\oplus$ is negative, reflecting the small overlap of the  DEM soft particles.  But $\Delta r_\oplus$ becomes less negative as the force balance changes for $t>0.05$~s.  When the sliding limit is reached, $\Delta r_\oplus$ decreases quickly to zero as the fine particle separates from the large particles (see video v2 in supplementary materials). 

\begin{figure}[H]
    \centering
    \includegraphics[width=0.5\linewidth]{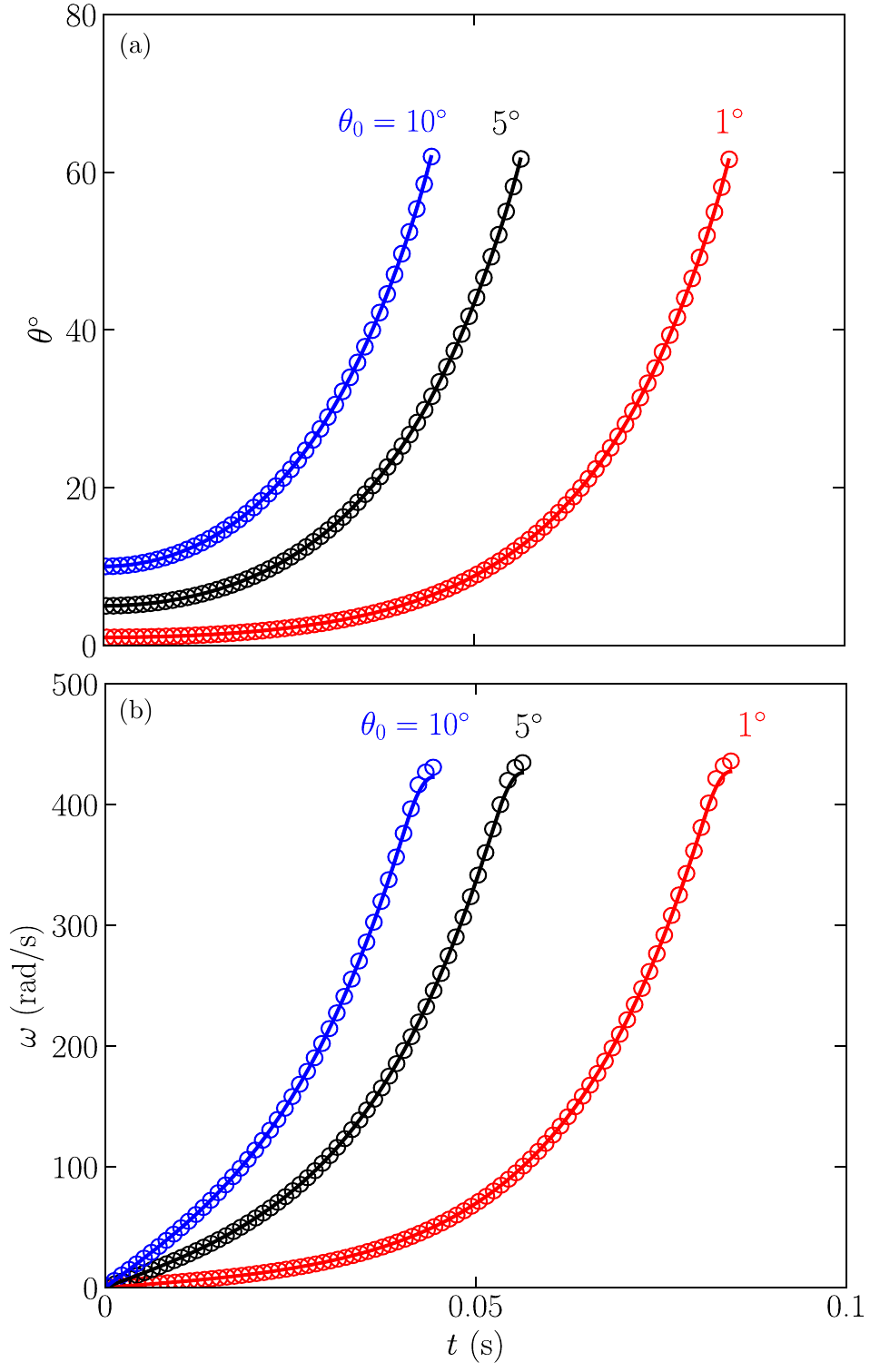}
    \caption{Comparison of DEM predictions of $\theta$ and $\omega$ (circles) for a fine particle with analytical solutions (solid curves) and Figure~\ref{fig:sch} geometry for different values of $\theta_0$ using the improved velocity-Verlet algorithm  ($\mathcal{R}=7$, $\mu=0.6$).}
    \label{fig:new}
\end{figure}

To verify the accuracy of this revised approach, we compare the DEM predictions for $\theta$ and $\omega$ with the analytical solution, as shown in Figure~\ref{fig:new} ($\mathcal{R}=7$, $\mu=0.6$). A close match between the predictions from the new approach and the analytical solutions, particularly the fact that the fine particle loses contact with the two large particles at $\theta \approx 61^\circ$, indicates that the revised approach provides accurate and physically realistic results.  In addition, the improved velocity-Verlet method is used to simulate the 10,100 cases included in Figure~\ref{fig:musr}. In every case the fine particle falls off as it should.

The improved velocity-Verlet integration also produces physically correct results for the limited number of cases we considered when rolling friction or a twisting force are included in the simulation. However, a detailed investigation of these effects is beyond the scope of this work.

% \begin{figure}[t]
%     \centering
%     \includegraphics[width=0.65\linewidth]{./figs/3ball/th1u/v3c.pdf}
%     \caption{Effect of modifications on the components of (a) $\mathbf{v}_{t,r}$ and (b) $\bm{h}$ in the $\hat{\mathbf{n}}_3$ direction and (c) radial distance of the fine particle from the line connecting the centers of the two large particles $\Delta r_{\oplus}$. FIGURE NOT USED IN TEXT. ITS ONLY FOR US TO SEE THE EFFECT OF THE NEW APPROACH. WE WILL REMOVE THIS IN THE FINAL DRAFT.}
%     \label{fig:ffinenew}
% \end{figure}

In Figure~\ref{fig:new}, note that the angular velocity at the end of the contact predicted by DEM is slightly higher than the analytical values for all three values of $\theta_0$. This difference arises because the analytical solution assumes a rigid sphere, whereas DEM simulations employ a soft sphere approach\footnote{The soft sphere approach causes particle overlap, altering $r_\oplus$ and generating a radial tangential force.}. Additional tests for small size ratio two-body oblique collisions show that the improved velocity-Verlet approach results match those of the standard approach, indicating that the improved approach maintains accuracy for small size ratio two-body problems. 

In methods such as Dissipative Particle Dynamics (DPD) and Smooth Particle Hydrodynamics (SPH), one approach to mitigate inaccuracy due to velocity-position lag is to use a projected velocity~\cite{grootDissipativeParticleDynamics1997,nikunenHowWouldYou2003}, typically calculated using Euler integration, which is only first-order accurate. This projected velocity is then used to compute inter-particle forces. While this approach may be adequate for DPD and SPH, it presents a significant challenge in DEM due to the integral in the history term (see Eq.\ref{eq:history}), where the accumulation of errors from Euler integration can lead to considerable inaccuracies. In contrast, the half-step velocity-Verlet method implemented in our work overcomes this limitation by achieving second-order accuracy in the history terms, as demonstrated in Eq.~\ref{eq:history2}.

\section{Fine particle percolation in static beds}

To highlight the significance of using the improved approach in practical problems with large size ratios, we simulate the percolation of fine particles through a static bed of large particles. This example captures an ensemble of initial and contact conditions for fine-large particle interactions. In addition to implementing the improved velocity-Verlet algorithm, careful selection of  DEM parameters, such as spring stiffness and timestep, is required for accurate simulations, particularly in granular systems with large size ratios. Thus, we begin by providing guidelines for selecting the appropriate DEM parameters, followed by a demonstration of the impact of using the revised integration scheme compared to the original one.

\subsection{Guidelines for selecting parameters in large size ratio DEM simulations}
\label{sec:guidelines}

When the size ratio is large, DEM simulation parameters need to be carefully selected. This becomes evident when considering the overlap.  While a particle overlap of 1\% of particle diameter might be typical in a simulation of similar size particles, that 1\% overlap for a large particle diameter would be a problematic 10\% overlap for a fine particle for $\mathcal{R}=10$. 

Consider first the normal spring stiffness, $k_n$, for a linear spring model. If $k_n$ is too large, it considerably increases the computational time because the timestep must be smaller. Conversely, if $k_n$ is too small, overlaps for fine particles can be too large in cases of large size ratios, potentially causing a fine particle to become embedded within a larger particle, resulting in unphysical outcomes. Therefore, selecting an appropriate value of $k_n$ is crucial.
One approach bases the spring stiffness on material properties and matches the contact duration in a linear spring model with the Hertzian contact duration~\cite{thorntonInvestigationComparativeBehaviour2011c}. While this method is effective for small size ratios where contacts are primarily driven by normal contact forces, it may be unsuitable for larger size ratios where effects such as sliding and rolling become more significant. 

We propose an alternative approach that determines $k_n$ based on the maximum tolerable overlap. 
For a linear spring model, the normal spring stiffness $k_n$ is related to the maximum overlap $\delta$ as
\begin{equation}
     k_n = \frac{m^* V_0^2}{\delta^2},
     \label{eq:kn}
\end{equation}
where $m^*$ is the effective mass, $V_0$ is a characteristic relative velocity between particles, and $\delta$ is the particle overlap. %For a given total mass and with an estimate of the characteristic relative velocity, we can choose $k_n$ such that the overlaps are small. 
For example, in the case of fine particles percolating through a static bed of large particles of the same material, which is shown in Figure~\ref{fig:bed}, $V_0^2 \approx 2gR$, where $g$ is the acceleration due to gravity and $R$ is the radius of the large particles. 
In this example, the large particle diameter is $2R = 4$ mm, $\mathcal{R}=$20, and the density for both large and fine particles is 2500 kg/m$^{3}$.
To ensure that the overlap between the large and fine particles remains small, say less than 0.1\% of the fine particle's diameter $d$, the minimum normal spring stiffness is 10271 N/m based on Eq.~\ref{eq:kn}.  %
The tangential spring stiffness is calculated using Eq.~\ref{eq:johnson}, which simplifies to Eq.~\ref{eq:johnson2} for particles of the same material. For $\nu = 0.3$,  $k_t = 8458$ N/m. Contact duration calculated using Eq.~\ref{tc} is $t_c =$ 3.2 \textmu s, indicating a timestep size $t_c/40 =$ 0.08 \textmu s.

A similar approach can be used for the Hertzian model based on the Young's modulus and Poisson's ratio~\cite{johnsonContactMechanics1985b}:
\begin{equation}
    \delta = \left(\frac{15 m^* V_0^2}{16 R^{*1/2} E^*}\right)^{2/5}.
    \label{eq:hertzdelta}
\end{equation}
To ensure an overlap of 0.1\% of the fine particle's diameter for our test case, the value of $E^*$ is 2.2 GPa based on Eq.~\ref{eq:hertzdelta}. So we can use $E = $ 4 GPa and $\nu = 0.3$. The tangential spring stiffness for the \textit{mindlin} model in LAMMPS is calculated as $k_t = 8G^*$~\cite{mindlinComplianceElasticBodies2021}, where $ G^*$ is the effective shear modulus: 
\begin{equation}
    \frac{1}{G^*} = \frac{2(2-\nu_l)(1+\nu_l)}{E_l}+\frac{2(2-\nu_f)(1+\nu_f)}{E_f}.
\end{equation}
The elastic contact time can be approximated as~\cite{johnsonContactMechanics1985b}
\begin{equation}  
    t_c \approx \frac{2.94 \delta}{V_0}.
\end{equation}
For the test case considered here, $t_c \approx 2.96$\,\textmu s. The computational timestep must therefore be smaller than $t_c/40 =0.074$ \textmu s.

Having selected these parameters, the next step is to choose the appropriate damping model in LAMMPS. To incorporate inelasticity LAMMPS traditionally requires a normal damping coefficient, $\eta_n$, as input. This value is calculated based on the desired coefficient of restitution, $e_n$, and a mean particle diameter, with the same $\eta_n$ value applied to all interactions of a given type (small-small, small-large, or large-large). For mixtures with a small range of particle sizes, this approach generally reproduces the desired restitution coefficient with minor variations. However, for mixtures with a large range of particle sizes, using a value for $\eta_n$ based on the mean particle diameter can result in inaccurate $e_n$ values that may deviate substantially from the value of $e_n$ specified for calculating the damping coefficient.

To address these issues in both the linear spring model and the Hertzian model, we have recently added a new damping model to LAMMPS named \texttt{coeff\_restitution}, which implements the approach outlined by Brilliantov et al.~\cite{brilliantovIncreasingTemperatureCooling2018} and is available in other open-source packages such as LIGGGHTS~\cite{klossModelsAlgorithmsValidation2012} and MercuryDPM \cite{weinhartFastFlexibleParticle2020}. 
It requires only the coefficient of restitution, $e_n$, as input. Once $e_n$ is specified, the model calculates the appropriate pair-wise damping coefficients based on whether the spring is linear or Hertzian, effectively accounting for the effects of particle size and ensuring that the specified $e_n$ is reproduced for each particle pair. Specifically, for a linear spring, the normal damping coefficient is 
\begin{equation}
    {\eta_n} = \sqrt{\frac{4m^* k_n}{1+\left( \frac{\pi}{\log(e_n)}\right)^2}},
\end{equation}
and for a Hertzian spring, the normal damping coefficient is calculated at each timestep using
\begin{equation}
    \eta_n =  -2\sqrt{\frac{5}{6}}\frac{\log(e_n)}{\sqrt{\pi^2+(\log(e_n))^2}}(R^* \delta)^{\frac{1}{4}}\sqrt{\frac{3}{2} \left(\frac{4}{3} E^*\right) m^*}.
\end{equation}
Note that LAMMPS also offers the \texttt{tsuji} model~\cite{tsujiLagrangianNumericalSimulation1992}, which also takes $e_n$ directly as an input. However, through simple two-sphere collision tests, we find that, at the time of testing, the \texttt{tsuji} model fails to accurately reproduce the specified $e_n$ values. 

The tangential damping coefficient $\eta_t$ is calculated by scaling the normal damping coefficient $\eta_t= \zeta \eta_n$. The scaling factor is $\zeta = \sqrt{k_t/k_n}$ for the linear spring model and $\zeta = \sqrt{4G^*/E^*}$ for the Hertzian model~\cite{thorntonInvestigationComparativeBehaviour2013b}.

Most DEM codes also provide models to incorporate rolling and twisting friction. LAMMPS includes a spring-dashpot-slider  \texttt{sds} model, which employs the pseudo-force formulation suggested by Luding~\cite{ludingCohesiveFrictionalPowders2008} to simulate rolling and twisting friction. Additionally, LAMMPS offers the \texttt{marshall} twisting friction model~\cite{MARSHALL20091541}, which is more suitable for Hertzian contacts. In this model, the necessary variables for twisting friction are calculated internally by the model, rather than relying on user-specified input values.

\subsection{Effect of using improved velocity-Verlet algorithm}
\label{sec:testcase}

\begin{figure}[h]
    \centering
    \includegraphics[width=0.3\linewidth]{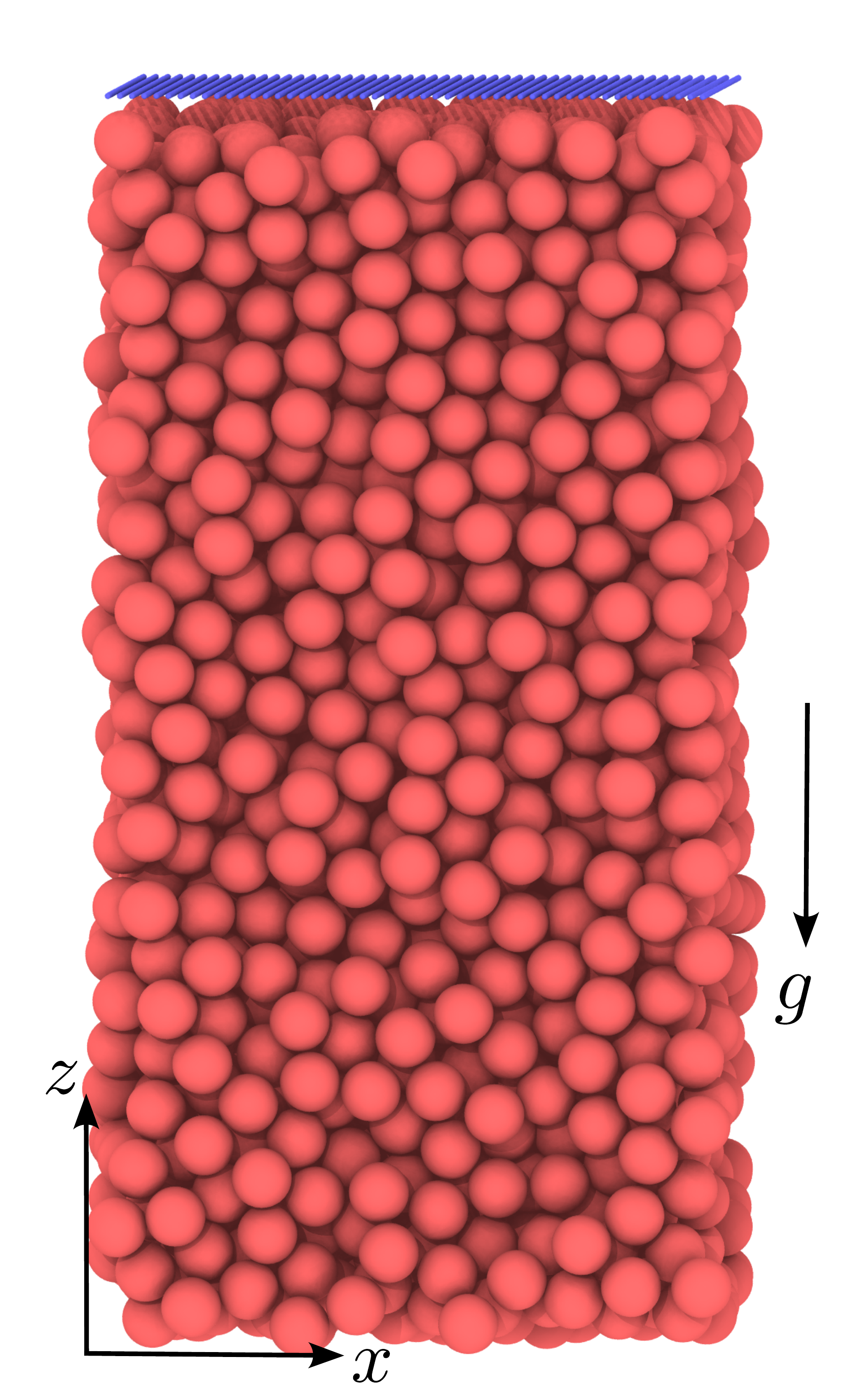}
    \caption{A static bed of large particles (red) with $R = 2$ mm. Fine particles (blue), $\mathcal{R}=20$, fall from just above the bed and percolate downward due to gravity.  The density for both large and fine particles is 2500 kg/m$^3$, $e_n = 0.8$, and $\mu = 0.3$.}
    \label{fig:bed}
\end{figure}

To compare the standard velocity-Verlet algorithm and the new approach outlined in this paper, we follow the above guidelines for parameter selection and simulate the situation in Figure~\ref{fig:bed}, applying the linear spring model in both cases.
A randomly packed bed of large particles, with an overlap of less than 0.1\%, is prepared using a standard particle growth algorithm \cite{lubachevskyGeometricPropertiesRandom1990a}. The monodisperse large particles have radius $R = 2$\;mm, and the density of both large and fine particles is 2500 kg/m$^3$. The static  bed of frozen large particles has dimensions of $20R \times 20R \times 40R$ and is periodic in the $x$ and $y$ directions, as shown in Figure \ref{fig:bed}. Non-interacting fine particles with $\mathcal{R}=20$ fall from 0.2 mm above the bed and percolate downward due to gravity. The fine particles interact with the bed particles but not with each other. The gravitational acceleration, $g$, is 9.81 m/s$^2$, the coefficient of restitution, $e_n$, is 0.8, and $\mu = 0.3$. The fundamental difference between this test case and the setup described in Figure~\ref{fig:sch} is that here the fine particle is not initially in tangential contact with the large particles. Instead, it comes into contact with the bed of large particles at varying velocities and angles. This allows us to model a wide spectrum of initial conditions for a fine particle contacting one or more large particles.

\begin{figure*}[t]
    \centering
    \includegraphics[width=0.95\linewidth]{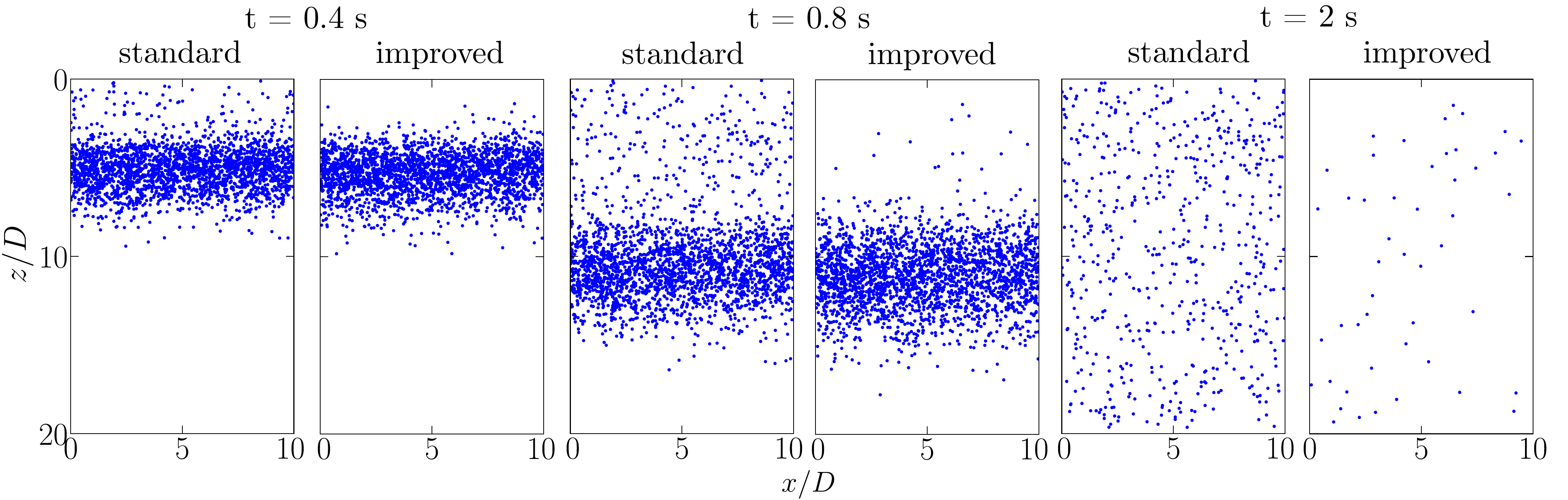}
    \caption{Fine-particle positions at different times (t = 0.4, 0.8, 2 s) for $\mathcal{R}$=20 and $\mu=0.3$, comparing standard and improved velocity-Verlet algorithms. Note that trapped particles in the improved approach represent valid solutions of the soft particle contact model (see Eq.~\ref{eq:mutana}) unlike most of the trapped particles in the standard case which do not.}
    \label{fig:perc}
\end{figure*}

We consider two cases where all parameters and other conditions are identical, including the static large-particle bed. The sole difference between them is the implementation of the velocity-Verlet algorithm. In the first case, we use the original version of the algorithm, referred to as the standard approach. In the second case, we employ the improved approach.
The locations of 2500 non-interacting fine particles at three different times after their release is shown in Figure~\ref{fig:perc}, where the bed particles are made transparent to better visualize the positions of the fine particles. 

When using the standard approach, many fine particles become unphysically trapped, as shown in Figure~\ref{fig:perc}, even though at  $\mathcal{R} = 20$ fine particles should predominantly percolate freely through the bed because fine particles are much smaller than pore throats in the bed~\cite{vyasImpactsPackedBed}. These trapped particles exhibit pendular motion similar to that shown in Figure~\ref{fig:3d2}, preventing them from percolating through the bed. Some fine particles become trapped within $z/D < 10$ at $t = 0.8$~s. By $t = 2$ s, approximately 25\% of the fine particles are trapped in the static bed.
In contrast, with the improved velocity-Verlet approach, the majority of fine particles percolate freely, forming a dense band. This behavior aligns with the expected outcome for $\mathcal{R} = 20$. 

A small fraction of fine particles, less than 2\%, still become trapped even with the improved approach. 
However, unlike the erroneous behavior discussed in Section 2, this occurrence is consistent with the model's expected behavior for soft particle simulations as described in the context of Eq.~\ref{eq:mutana}. 
When $\mu > \tan{\alpha}$, fine particles can become trapped within the static bed of large particles when a fine particle with high velocity impacts and slides into a valley between large particles. In such cases, the large static friction generated during the initial contact prevents the fine particle from reaching the sliding limit, resulting in its entrapment. 
This is similar to a basketball getting stuck between the rim and the backboard. The softer the basketball, the more likely it gets stuck, although it still happens infrequently because the basketball needs to hit the rim and backboard in just the right way.
In granular beds composed of large particles, the random packing and associated gaps between neighboring large particles further increases the likelihood of such physical trapping as they correspond to smaller $\alpha$ values at a given $\mathcal{R}$ compared to the case where large particles are touching.
In the absence of twisting friction, these trapped particles exhibit pendular motion. However, unlike cases with the standard velocity-Verlet algorithm, they are not drawn deeper into the valley between large particles. While the pendular motion (in the absence of twisting friction) is physically unrealistic, the observed trapping is consistent with real-world physics for soft particles.

\section{Summary}

As computational techniques evolve and researchers apply DEM simulations are applied to new and complex domains, it is vital to proceed with caution, as demonstrated by the molecular dynamics code used in the supercooled water example~\cite{Smart2018} mentioned in the introduction. While expanding the application of DEM to more challenging problems is appealing, it also presents unforeseen challenges that can compromise the accuracy of simulations\footnote{Platforms like the Open Network on DEM Simulations (ON-DEM)~\cite{ondem} are valuable for tracking progress and ensuring consistency across open-source codes in the field as they evolve.}.
Our findings underscore the importance of rigorous validation and the need for careful consideration when extending established methods to novel research contexts and parameter regimes.

The improved velocity-Verlet integration scheme proposed in this paper addresses critical limitations in standard implementations of the DEM. We demonstrate that the conventional scheme can lead to unphysical results where fine particles remain in contact with larger particles due to the emergence of erroneous contact forces. This issue arises for all size ratios and friction coefficients, and produces unphysical trapping even at size ratios as low as three when the friction coefficient is high. The improved velocity-Verlet method resolves these inconsistencies by assuring that the velocity and unit normal vector are updated at the same  time, rather than a half timestep apart. Since the computational cost of algebraic operations is orders of magnitude smaller than more expensive tasks like neighbor detection and writing output data, the additional computational overhead related to the improved velocity-Verlet algorithm is negligible.

Given the complexities associated with simulating systems involving large particle size ratios, we also offer guidelines for selecting appropriate spring stiffness, timesteps, and damping parameters for high size ratio simulations. These guidelines ensure the robustness and stability of simulations, especially when dealing with highly size-polydisperse systems.
Simulations of fine particle percolation through a static bed of large particles demonstrates the usefulness of improved approaches in providing a more accurate and physically consistent description of particle behavior.
With recent advances in computational power enabling the simulation of granular systems with increasingly wider ranges of size-polydispersity that capture the complexity of real-world industrial and geophysical granular flows, it is necessary that simulations with size ratios $\mathcal{R} \geq 3$ utilize the improved velocity-Verlet method to maintain accuracy and avoid artifacts that can compromise results.

\section*{Funding information}
This material is based upon work supported
by the National Science Foundation under
Grant No.\ CBET2203703.

\section*{Acknowledgments}

We extend our gratitude to Joel Clemmer and Dan Bolintineanu (Sandia National Laboratories) for their insightful discussions and assistance with LAMMPS. Our thanks also go to P. J. Zrelak (University of Oregon) and Eric Breard (University of Edinburgh) for their invaluable help with MFiX and MFiX-Exa. We are grateful to Anthony Thornton (University of Twente) for his support with MercuryDPM, and to Stefan Luding (University of Twente) and Ken Kamrin (Massachusetts Institute of Technology) for their enlightening discussions, which greatly enhanced our understanding of this problem.

\section*{Declaration of AI-Assisted Tools in the Writing Process}

During the preparation of this work, the authors utilized ChatGPT for improving writing and proofing. After employing this tool/service, the authors reviewed and edited the content as necessary and assume full responsibility for the content of the published article.

\section*{Appendix 1: Analytical Solution}
\label{app1}
\begin{figure*}[h]
    \centering
    \includegraphics[width=0.8\linewidth]{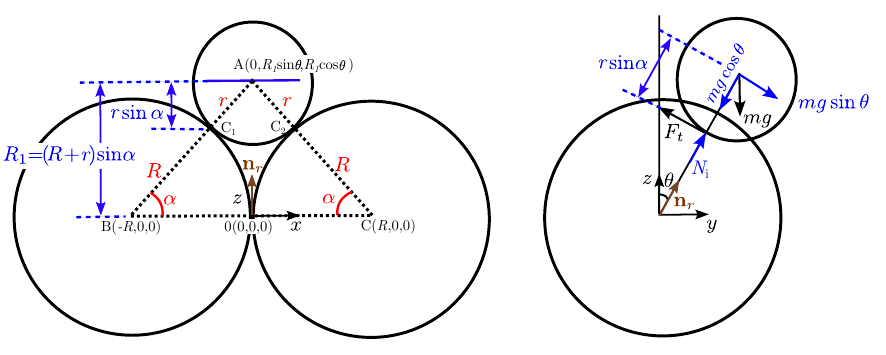}
    \caption{Schematics showing (left) front view of the three particle geometry and (right) side view of forces on the fine particle.}
    \label{fig:anaschem}
\end{figure*}

The geometry of the problem is shown in Figure~\ref{fig:anaschem}. Initially, and in the rigid body limit, the fine particle of mass $m$ and radius $r$ moves in a state of pure rolling. Once the tangential friction force $F_t$ equals $\mu F_n$, the particle starts to slide in addition to rolling. 

For initial pure rolling, the velocity of the fine particle, 
$v_\theta$, can be determined by equating kinetic and gravitational potential energy. The gain in kinetic energy for an incremental change in the angle $\delta \theta$ of a particle undergoing pure rolling is
\begin{equation}
    \Delta KE = \frac{1}{2}m v_{\theta+\delta \theta}^2 - \frac{1}{2}m v_{\theta}^2 + \frac{1}{2}I \omega_{\theta+\delta \theta}^2 - \frac{1}{2}I \omega_{\theta}^2,
    \label{eq:kegain}
\end{equation}
where $I=2 m r^2/5$ is the moment of inertia of the fine particle and its the angular velocity is
\begin{equation}
    \omega_\theta = \frac{v_\theta}{r \sin{\alpha}},
\end{equation}
where
\begin{equation}
    \alpha = \cos^{-1}\left(\frac{R}{r+R}\right) = \cos^{-1}\left(\frac{\mathcal{R}}{1+\mathcal{R}}\right).
    \label{eq:alpha}
\end{equation}
For $\delta \theta \to 0$, this simplifies to
\begin{equation}
    \Delta KE = \left(\frac{5\sin^2\alpha+2}{5\sin^2\alpha}\right) v_\theta m \frac{dv_\theta}{d\theta}\delta\theta.
\end{equation}
For this incremental change $\delta \theta$, the change in potential energy
\begin{equation}
    \Delta PE = mg(R+r)\sin{\theta}\sin{\alpha}\delta\theta.
\end{equation}
Equating the $\Delta PE$ and $\Delta KE$ gives
\begin{equation}
    v_\theta \frac{dv_\theta}{d\theta} = \frac{5\sin^2\alpha}{5\sin^2\alpha+2} g(R+r)\sin{\theta}\sin{\alpha}.
    \label{eq:vthera}
\end{equation}
Integrating Eq.~\ref{eq:vthera} for the case where the velocity at the initial release angle $\theta_0$, $v_\theta(\theta_0) = 0$, results in
\begin{equation}
    v_{\theta} =(R+r)\sin\alpha\frac{d\theta}{dt} = \left[\frac{10g(R+r)\sin^3\alpha}{5\sin^2\alpha+2}(\cos{\theta_0}-\cos{\theta})\right]^{1/2}.
    \label{eq:anaroll}
\end{equation}
The tangential force is 
\begin{equation}
    F_t = \frac{I\dot{\omega}}{r\sin\alpha},
    \label{eq:ft1}
\end{equation}
where the angular acceleration of the fine particle is 
\begin{equation}
    \dot{\omega} = \frac{d\omega}{dt} = \frac{1}{r\sin\alpha}\frac{dv_{\theta}}{dt}.
    \label{eq:acc}
\end{equation}
Note that both Equations~\ref{eq:ft1} and~\ref{eq:acc} use the projected radius $r\sin{\alpha}$ instead of $r$. This is because, as shown in Figure~\ref{fig:anaschem}, the normal distance between the fine particle center and the contact points is $r\sin{\alpha}$.
The total tangential force is evenly split between the two contact points C$_1$ and C$_2$ as
\begin{equation}
    F_{t_1}=F_{t_2}=\frac{F_{t}}{2}.
\end{equation}
The net normal force is
\begin{equation}
    N = mg\cos\theta - m(R+r)\sin\alpha\left(\frac{d\theta}{dt}\right)^2,
\end{equation}
which is also split between the two contacts as
\begin{equation}
    N_{1}=N_{2}=\frac{N}{2\sin\alpha}.
\end{equation}
%Implementing Coulomb's limit
%\begin{equation}
 %   \textit{ if }F_{t_i}>\mu N_i,\textit{ then }F_{t_i}=\mu N_i.
%\end{equation}
%takes us to the second stage of motion which consists of sliding where

When the fine particle begins to slide ($F_{t_i}>\mu N_i$), following Newton's second law, $v_\theta$ then evolves as 
\begin{equation}
    \frac{dv_\theta}{dt}=g\sin\theta-\frac{\mu N}{m\sin{\alpha}}. 
    \label{eq:anaslide}
\end{equation}

\bibliographystyle{elsarticle-num} 
\bibliography{dem_refs}

\end{document}